\documentclass[useAMS,usenatbib]{mn2e}
\usepackage{cleveref}
\usepackage{natbib}
\usepackage{graphicx}
\usepackage{amsmath}

\newcommand{\msun}{M$_\odot$}
\newcommand{\lsun}{L$_\odot$}
\newcommand{\rsun}{R$_\odot$}
\newcommand{\kms}{km~s$^{-1}$}
\newcommand{\kepler}{{\it Kepler }}
\newcommand{\Teff}{\mbox{$T_\mathrm{eff}$}}
\newcommand{\Teffw}[1]{\mbox{$\Teff\hspace{-0.5mm} =\hspace{-0.5mm} #1 \,\mathrm{kK}$}}
\newcommand{\Msol}{$M_\odot$}

\newcommand{\Ion}[2]{\ion{#1}{#2}}
\newcommand{\Ionw}[3]{\mbox{\ion{#1}{#2}~$\lambda\,#3$\,\AA}}
\newcommand{\Ionww}[3]{\mbox{\ion{#1}{#2}~$\lambda\lambda\,#3$}}
\newcommand{\logg}{\mbox{$\log g$}}
\newcommand{\loggw}[1]{\mbox{$\log g\hspace{-0.5mm} =\hspace{-0.5mm}  #1$}}
\newcommand\ion[2]{#1$\;${\scshape{#2}}}

\def\lesssim{\mathrel{\hbox{\rlap{\hbox{\lower3pt\hbox{$\sim$}}}\hbox{\raise2pt\hbox{$<$}}}}}
\def\gtrsim{\mathrel{\hbox{\rlap{\hbox{\lower3pt\hbox{$\sim$}}}\hbox{\raise2pt\hbox{$>$}}}}}

\title[Identifying close binary central stars of PN with \kepler]{Identifying close binary central stars of PN with \kepler}
\author[De Marco et al.]{Orsola De Marco$^{1,2}$\thanks{E-mail:
orsola.demarco@mq.edu.au} , J. Long$^3$, George H. Jacoby$^3$, T. Hillwig$^4$, M. Kronberger$^5$, \newauthor Steve B. Howell$^{6}$, N. Reindl$^{7,8}$ and Steve Margheim$^9$\\
$^{1}$Department of Physics and Astronomy, Macquarie University, Sydney NSW 2109, Australia\\
$^{2}$Astronomy, Astrophysics and Astrophotonics Research Centre, Macquarie University, Sydney NSW 2109, Australia\\
$^{3}$Giant Magellan Telescope / Carnegie Observatories, Pasadena, CA 91101, USA\\
$^{4}$Valparaiso University, 700 Chapel Dr, Valparaiso, IN 46383, USA\\
$^{5}$Deepskyhunters Collaboration\\
$^{6}$NASA Ames Research Centre Moffett Field, CA 94035, USA\\
$^{7}$T\"ubingen University, Geschwister-Scholl-Platz 72074 T\"ubingen, Germany\\
$^{8}$Institute for Astronomy and Astrophysics, Kepler Center for Astro and Particle Physics, Eberhard Karls University, 72076, T\"ubingen, Germany\\
$^9$Gemini Observatory, Southern Operations Center, Casilla 603, La Serena, Chile}
\begin{document}

\date{Received; Accepted}

\pagerange{\pageref{firstpage}--\pageref{lastpage}} \pubyear{2014}

\maketitle

\label{firstpage}

\begin{abstract}
Six planetary nebulae (PN) are known in the \kepler space telescope field of view, three newly identified.
Of the 5 central stars of PN with useful \kepler data, one, J~193110888+4324577, is a short-period, post common envelope binary exhibiting relativistic beaming effects. A second, the central star of the newly identified PN Pa~5, has a rare O(He) spectral type and a periodic variability consistent with an evolved companion, where the orbital axis is almost aligned with the line of sight. The third PN, NGC~6826 has a fast rotating central star, something that can only be achieved in a merger.  Fourth, the central star of the newly identified PN Kn~61, has a PG1159 spectral type and a mysterious semi-periodic light variability which we conjecture to be related to the interplay of binarity with a stellar wind. Finally, the central star of the circular PN A~61 does not appear to have a photometric variability above 2 mmag. With the possible exception of the variability of Kn~61, all other variability behaviour, whether due to binarity or not, would not easily have been detected from the ground. We conclude, based on very low numbers, that there may be many more close binary or close binary products to be discovered with ultra-high precision photometry. With a larger number of high precision photometric observations we will be able to determine how much higher than  the currently known 15 per cent,  the short period binary fraction for central stars of PN is likely to be.
\end{abstract}


\section{Introduction}
\label{sec:introduction}

What is a planetary nebula? The textbooks are clear and definitive that a planetary nebula (PN) is a common, short-lived, phase of a typical star like the Sun that occurs as it burns up the last of its available nuclear fuel. During the next 1000 years or so, the star blows off its outer layers to reveal a central core that becomes hot enough to ionise the ejected layers. That ionised gas then glows due to recombination and collisional excitation. This simple scenario likely has some validity. However, we still do not have a satisfactory physical explanation for how single stars create PN with non-spherical shapes, which amount to 80-85 percent of the entire population \citep{Parker2006}.

%but highly compelling observational evidence has been building over the past 10-15 years to force a re-examination of that basic model in terms of binary star effects \citep[for a review see][]{DeMarco2009}.

%One expects mass-loss from a star to be spherically symmetric, and consequently, the nebula around a remnant central star is expected to look like a spherical shell, such as, for instance, Abell~39 or Pa~9 \citep{Jacoby2001,Jacoby2010}. Spherical PN, however, represent only 15-20 per cent of the PN population \citep{Parker2006}. Bipolarity has observationally been linked to binary progenitors \citep{Miszalski2009} and axi-symmetry or point-symmetry are also explained theoretically most easily by a binary shaping process \citep{Soker1997,Mastrodemos1999}. Thus one may expect, circumstantially, that $\sim$80-85 per cent of all PNe have experienced an interacting binary phase in their late evolution. 

\citet{GarciaSegura1999,GarciaSegura2005} argued that stellar rotation and magnetic fields in {\it single} stars can alter the morphology of a mass-losing asymptotic giant branch (AGB) star, but \citet{Soker2006} and \citet{Nordhaus2007} argued that in a large majority of cases, single AGB stars are unlikely to be able to sustain large-scale magnetic fields for long enough to affect shaping, because the field embedded in an expanding plasma, drains the star of angular momentum on short time scales and quenches itself. Without an angular momentum source, the field would vanish. In the models of \citet[][]{GarciaSegura1999} and \citet[][]{GarciaSegura2005} the field strength was assumed constant and was not coupled to the negative feedback action of the stellar envelope. In fact, recently \citet{GarciaSegura2014} have re-examined this issue and have determined, using stellar structure and evolution simulations, including the effect of angular momentum transport, that AGB stars cannot maintain rotation profiles that allow the formation of bipolar PN. A stellar or substellar companion, though, can provide the angular momentum source \citep{Soker2006,Nordhaus2007}. 
%If this were what actually happens, then PNe would be, by and large, a binary interaction phenomenon.

A host of related conundrums that also may find explanations in the presence of a binary companion, are the unexplained intensifying of mass-loss at the end of the AGB \citep[but see][]{Lagadec2008}, linear momenta in post-AGB outflows 1000 times in excess of what can be explained by radiation pressure on dust \citep{Bujarrabal2001}, or the lack of an explanation for the constant bright edge of the PNLF in old elliptical and young spiral populations \citep[][ but see also \citet{Schoenberner2007}]{Jacoby1992,Marigo2004}. Finally, the population synthesis studies of \citet{Moe2006} and Moe \& De~Marco (2010) show that the current scenario predicts too many PN in the Galaxy by a factor of 5 at the 3$\sigma$ level, while the binary interactions (from strong, such as common envelope [CE] interactions to weak, such as gravitational focussing) can explain $\sim$70 per cent of all PN.

The {\it binary hypothesis} \citep{DeMarco2009b} therefore argues that some single AGB stars do not make a PN, which implies that binarity is over-represented in the PN population compared to the progenitor main sequence population. To test this emerging hypothesis, several groups are actively working together to determine the true frequency of binary central stars and the properties of those systems \citep{Bond2000,DeMarco2004,Miszalski2009,DeMarco2013,Hillwig2010,Jones2014}. 

The only binary fraction that is reasonably well constrained today is that of very close central star binaries: $\sim$15 per cent of all central stars are in post-CE binaries (periods typically shorter than a day) and are detected with periodic flux variability due to irradiation, ellipsoidal variability, or eclipses \citep{Bond2000,Miszalski2009}. Almost all of the photometric variables detected have amplitudes larger than $\sim$0.1~mag, mostly due to suboptimal sampling due to telescope scheduling and weather constraints. 

It has therefore been asked how many objects have remained undetected as one would expect that inclination, companion size and orbital distance would all play a role in the variability amplitude \citep{DeMarco2008c}.  
It is likely that there is a dearth of central star close binaries with periods just longer than those detected so far (i.e., between a couple of days and a couple of months). It is likely that with our current detection limits we would have easily detected binaries with periods up to two weeks \citep{DeMarco2008c}. In addition, studies of post-CE white dwarf (WD) binaries, the progeny of post-CE central stars of PN binaries, seem to indicate a dearth of objects in that period range \citep{Schreiber2009}. However, short period, post-CE binaries with smaller companions, with an unfavourable orbital inclination, or with a weak irradiation effect should  be there, with variability below the ground-based threshold.
We have therefore embarked in a study to find new post-CE central star binaries with \kepler to determine whether we can get an indication of the frequency of the undetected, low amplitude binaries.

While a sample of 6 central stars in the \kepler\ field precludes the statistical  revision of the short-period binary fraction, it would be indicative to find even just one photometric variable that would {\it not} have been detected from the ground. Any indication that such binaries exist would motivate further studies to quantify how much larger the post-CE binary fraction is than the 15 per cent currently known. 
%This would be particularly important considering that the current fraction is already 2-3 times higher than the few percent that are predicted from the main sequence binary fraction and period distribution \citep{Raghavan2010}. 
%A larger post-CE central star binary fraction would strongly indicate that CE interaction is a preferential channel in the formation of PN.

In \S~\ref{sec:sample} we present the PN sample in the \kepler field of view, while in \S~\ref{sec:observations} we describe the observations. In \S~\ref{sec:j19311} to \S~\ref{sec:pa5} we present the \kepler lightcurves, atmosphere model analyses, as well as combined analysis of the light and radial velocity curves to derive stellar and binary parameters for five of the six central stars of PN in the {\it Kepler} field of view that have adequate data. We conclude in \S~\ref{sec:conclusions}.

\section{The PN Sample}
\label{sec:sample}

\begin{figure}
\centering
\includegraphics[scale= 0.45]{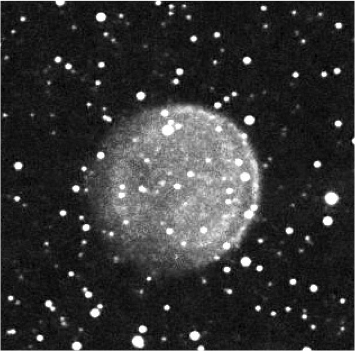}
\includegraphics[scale= 0.45]{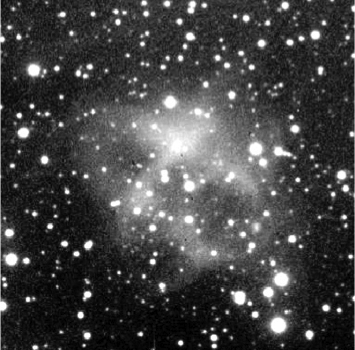}
\includegraphics[scale= 0.45]{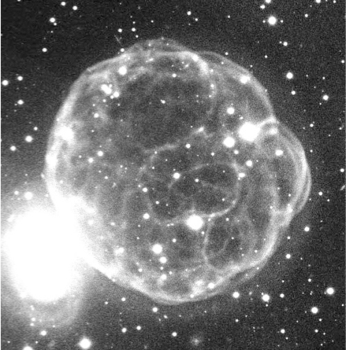}\\
\includegraphics[scale= 0.45]{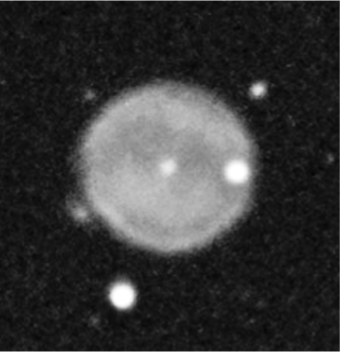}
\includegraphics[scale= 0.45]{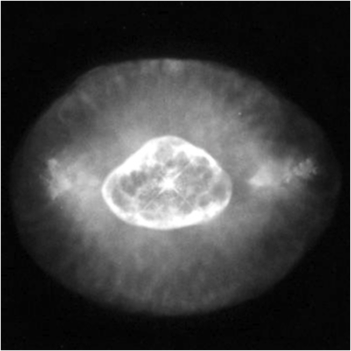}
\includegraphics[scale= 0.45]{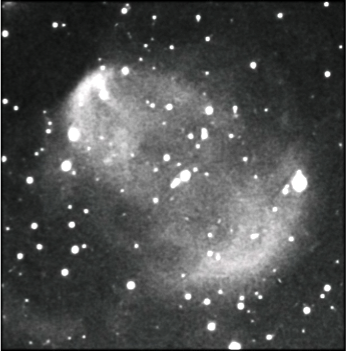}
\caption{A mosaic of the six PN in the \kepler field that were observed by our programme (the nebular diameters can be found in Table~\ref{tab:data}; North is towards the top, East to the left, except for Kn~61 - see caption to Figure~2). Top left: A~61, courtesy of Jim Shuder; Top centre: J19311+4324, \citet{Aller2013}; Top right: Kn~61, courtesy of Travis Rector and the Gemini Observatory; Bottom left: NGC~6742, courtesy of Adam Block. Bottom centre: NGC~6826, courtesy of Bruce Balick; Bottom right: Pa~5, from \citet{Jacoby2010}}
\label{fig:images}
\end{figure}

We have monitored (programme \# GO30018) six PN in the \kepler field. Three were known (Abell~61, hereafter A~61, NGC~6742 and NGC~6826). One was recently discovered by \citet[][the central star is known as J193110888+4324577, hereafter J19311, while the nebula may be indicated as AMU~1 after the discoverers, but we shall use the star's name to indicate this object]{Aller2013}. Two were found by the {\it Deep Sky Hunter} team \citep[][Kronberger~61, hereafter Kn~61 and Patchick~5, hereafter Pa~5]{Jacoby2010}, partly during an effort targeting the \kepler field in support of this study. A seventh object, PaTe~1, also found by the Deep Sky Hunters, is in an area of extended emission but is not seen in [O III] images, and therefore cannot be confirmed as a PN at this time.  Available data for these objects are summarised in Table~\ref{tab:data} while images are presented in Fig.~\ref{fig:images}.

A~61 is an old, circular PN first presented by \citet{Abell1966}. Its central star spectrum was modelled by \citet{Napiwotzki1999} and the following parameters were derived: $T_{\rm eff} = (88\,000\pm7900)$~K, $\log g=7.10\pm0.37$, $R=0.67$~\rsun\ at a distance of 1380~pc. \citet{Drummond1980} lists the central star of A~61 as a variable and the Russian catalogue of variable stars \citep{Kukarkin1981} designates it as NSV~11917.

\kepler data for the central star J19311 were first presented by \citet{Ostensen2010}, who showed that it has a periodically variable light curve. The PN was only recently discovered by \citet{Aller2013} who showed it to consist of two elongated structures at 90~deg from one another and with kinematics indicating that they were ejected almost at the same time.

Kn~61 is a PN discovered by the  {\it Deep Sky Hunters}  \citep{Kronberger2012} and already studied by \citet{GonzalezBuitrago2014}, who showed it to have a hydrogen poor, PG1159-type central star. Its central star's spectrum was originally presented by \citet{Jacoby2012}, \citet{Long2013} and \citet{GarciaDiaz2014}. \citet{GarciaDiaz2014} discuss extensively the PN, concluding that it is hydrogen poor, a quality shared by only a handful of other known objects \citep{Harrington1997}.

NGC~6742 is a little studied, circular PN with a morphology which is very similar to that of A~61 (Fig.~\ref{fig:images}). Its central star's light is diluted in the bright nebula and in \S~\ref{sec:observations} we question whether the \kepler photometry could even identify the star itself.

NGC~6826 is a very well studied PN. Its central star was detected to vary by \citet{Bond1990b} and classified as a ZZ Leporii star by \citet{Handler2003b}. The stellar spectrum was modelled by \citet[][$T_{\rm eff}$=46~kK, $\log g$ = 3.8, consistent with a mass of 0.74~\msun]{Kudritzki2006}. The PN consisted of two, well defined, elliptical structures and two jet-like structures aligned with the long axis of the PN. The \kepler\ light curve of the central star was presented by \citet{Handler2013} who demonstrated convincingly that its variability behaviour is not consistent with a short period binary, but instead with a fast rotating central star. 

\begin{figure}
\centering
\includegraphics[scale= 0.25]{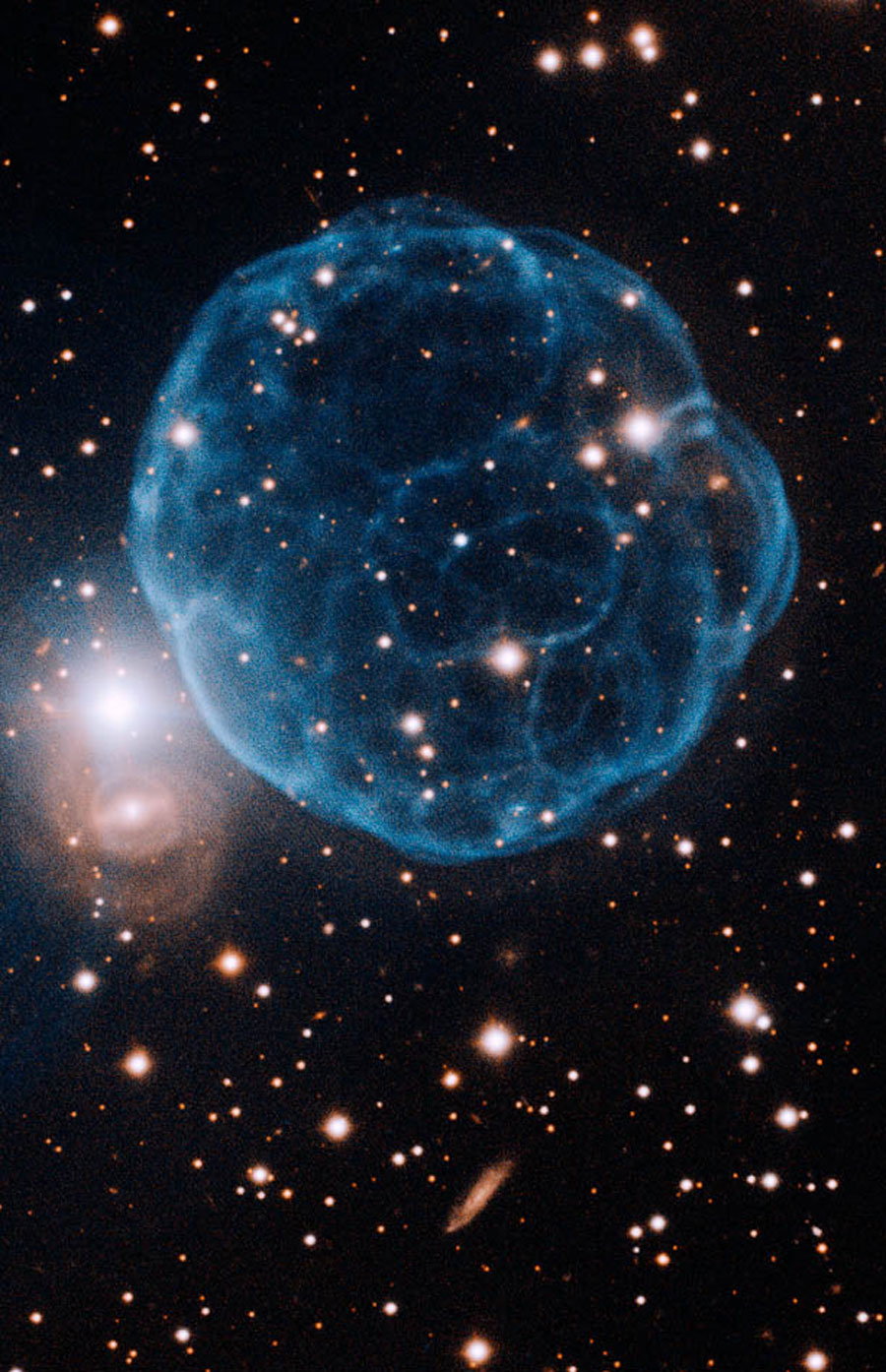}
\caption{A Gemini-north GMOS, 500-second exposure image of the newly-discovered PN Kn~61 obtained with the Gemini telescope. [OIII] is blue, H$\alpha$ is red. North is 21 degrees West of the vertical, West is 90 degrees left of North and the field of view is 2.2 x 4.4 arcminutes. Image curtesy of the Gemini Observatory and Travis Rector (see Gemini Office press release http://www.gemini.edu/node/11656)}
\label{fig:kn61-image}
\end{figure}

The  \kepler lightcurve of the central star of Pa~5 (also known as J19195+4445), was already presented by \citet{Ostensen2010}. The PN was discovered by the {\it Deep Sky Hunters} (DSH J1919.5+4445; \citealt{Kronberger2006}). \citet{Ostensen2010} also presented a spectrum of the central star and remarked on this star being a very hot PG1159 star, a classification that we will revise in \S~\ref{sec:pa5}. 
\begin{table*}
\begin{tabular}{lcccccc}
\hline
                                        &  A~61               &  J~19311        &  Kn~61        &  NGC~6742          &  NGC~6826  &  Pa~5\\
\hline
 \kepler Input Catalogue &  9583158      & 7755741        & 3231337       &     10963135      & 12071221         & 8619526 \\
 RA                                 &  19 19 10.22   & 19 31 08.89   & 19 21 38.93   &  18 59 20.03      & 19 44 48.15    & 19 19 30.52 \\
 Dec                               &  +46 14 52.03  & +43 24 57.74 &  +38 18 57.36 &  +48 27 55.24   & +50 31 30.26    & +44 45 43.08\\
 Distance (kpc)             &    1.38$^c$       &   1.82$^e$        &  4::                  & 4.86$^e$         &   1.3$^e$        &1.4 \\
 PN diameter (arcsec) &      200$^d$        & $\sim$300$^f$ &   104                &      27$^g$        &  27x24$^e$      &         120$^i$    \\
 PN morphology           & circular              & bipolar/elliptical  &  $\sim$circular  &   circular          &   elliptical/jets  &  bipolar/torus\\
 $V$-band magnitude$^a$ &17.389             & 13.697             & 18.416           & 16.567$^h$  & 10.41$^g$        & 15.670$^j$\\
\kepler magnitude       &17.321                &13.752                  &18.283                &16.255$^h$        & 10.757             &15.839\\
 Central star spectral type$^b$ &        DAO        & O(H)                     & PG1159          &  ?                       & O3f(H)             &  O(He) \\
\# of \kepler ``quarters"     & 10 & 7 & 4 & 1 & 7 & 8 \\
Light curve period (days)          &--           & 2.928        & 2-12                 & --     & 0.619,1.236      & 1.12 \\
Light curve amplitude (mmag)  & $<$2    & 0.729        & $\sim$80-140   & --     & $\sim$2-8             & 0.5   \\
\hline
\multicolumn{7}{l}{$^a$From \citet{Everett2012} unless otherwise stated; $^b$We stay away from sub-dwarf classifications for these objects, which confuses  }\\
\multicolumn{7}{l}{\ \ them with post-RGB stars, and use instead the classification system used by, e.g., \citet[][]{Mendez1988a}; $^c$Napiwotzki 1999; $^d$ESO  }\\
\multicolumn{7}{l}{\ \ catalogue; $^e$Frew (2008); $^f$This nebula was decomposed into two elliptical/bipolar shells approximately 5$\times$2~arcmin$^2$ with axes  }\\ 
\multicolumn{7}{l}{\ \ perpendicular \citep{Aller2013} to each other and with similar expansion velocities and kinematic ages; $^g$\citet{Acker1992}; }\\
\multicolumn{7}{l}{\ \  $^h$Likely too bright because of the bright nebula; $^i$Width of the torus' long axis; $^j$but see \S~\ref{ssec:kn61-spectrum} for an alternative value.}\\

\end{tabular}
\caption{Available literature data and results for the sample of PN in the \kepler field of view \label{tab:data}}
\end{table*}

%Distance N6826 is from expansion Frew 2008
% Distance to Kn61 assumed see Gonzalez-Buitrargo et al. APN6
% Pa5 = J18185+4445 = DSH J1919.5+4445 Kronberger et al. (2006)

In Fig.~\ref{fig:images} and \ref{fig:kn61-image} we present images of the six PN. The image of Kn~61 was obtained with the Gemini  telescope in July 2011 after a confirmation image was obtained by us at the KPNO 2.1-m telescope.

\section{Description of the Observations}
\label{sec:observations}

\subsection{{\it Kepler} photometry, periodicity analysis, and non-detections}
\label{ssec:keplerphotometry}

The {\it Kepler} space telescope was launched on March 7, 2009 with the aim of monitoring approximately 115 sq. degrees of sky centred on the Cygnus-Lyra region of the sky. The telescope aperture is 1.4 m. The filter bandpass has a full width at half maximum between 4300 and 8900 \AA. The monitoring cadence of our targets was approximately 30 minutes while each exposure time was equivalent to 1626 sec. The \kepler pixel scale is 3.98 arcsec/pix. The \kepler data is distributed by quarters, each three months long. The quarters available for a particular object vary: A~61 was observed in quarters \#2-7 and \#10-13, J19311 was observed in quarters \#1-6 and \#13, Kn 61 in quarters \#10-13, NGC 6742 was only observed for a single quarter (\#13). NGC 6826 has data from quarters \#1-7 and \#10-13 and Pa 5 was observed in quarters \#1, \#5-7, and \#10-13.

Although \kepler's detectors contain over 95 million pixels, only a small fraction can be captured for the creation of long cadence light curves due to spacecraft onboard storage and downlink constraints. As a result, targets chosen by observers are assigned a mask, and only pixels within the mask are stored and eventually downloaded for analysis. Light curves from \kepler are obtained through simple aperture photometry on the recorded pixels, and processed by the \kepler ``pre-search data conditioning" (PDC) pipeline to remove systematic errors. The same measures in the pipeline that try to minimise false positives for exoplanet transit detection also risk removing intrinsic star variability. However, comparing PDC light curves to ones extracted from the original target pixel files and detrended with the cotrending basis vectors provided with the data release showed them to be very similar, leading us to conclude that PDC light curves were acceptable for our purposes without additional reduction.

Each quarter of data was analysed to detect possible periodic components in the signal using the Lomb-Scargle periodogram algorithm. The resulting periodograms were used to identify the strongest periodic component in the light curve. In order to combine different quarters of data for an object, it was necessary to normalise the light curves by the median values. Otherwise, differences between quarters result in discontinuities when ÒstitchingÓ the light curves. A periodogram was also generated for the stitched light curve, to get a better signal to noise ratio in detecting the periodic signal. To generate the folded light curve, the stitched data was folded at the strongest period from the periodogram. In the phasing process, the data were binned to form median flux values, with 50 bins per light curve cycle.

The central star of A~61 has no detected variability to a limit of 2~mmag, which is curious given its designation as NSV~11917. 

For the central star of NGC~6742 we did not detect any variability. However, we were unable to distinguish any difference between the flux in pixels centred on the central star coordinates from pixels a few arcsec away. For the other PN, the nebulae have very low surface brightnesses relative to the central stars. In the case of NGC 6742, however, \kepler's large pixel size and broad bandpass results in a high background noise that degrades the contrast between the nebula and star. We therefore must conclude that this target could not measured accurately enough to draw any conclusion as to the variability of the stellar light.
%J19311 multi
\begin{figure*}
\centering
\includegraphics[scale= 0.35,clip=true,trim=0mm 70mm 0mm 0mm]{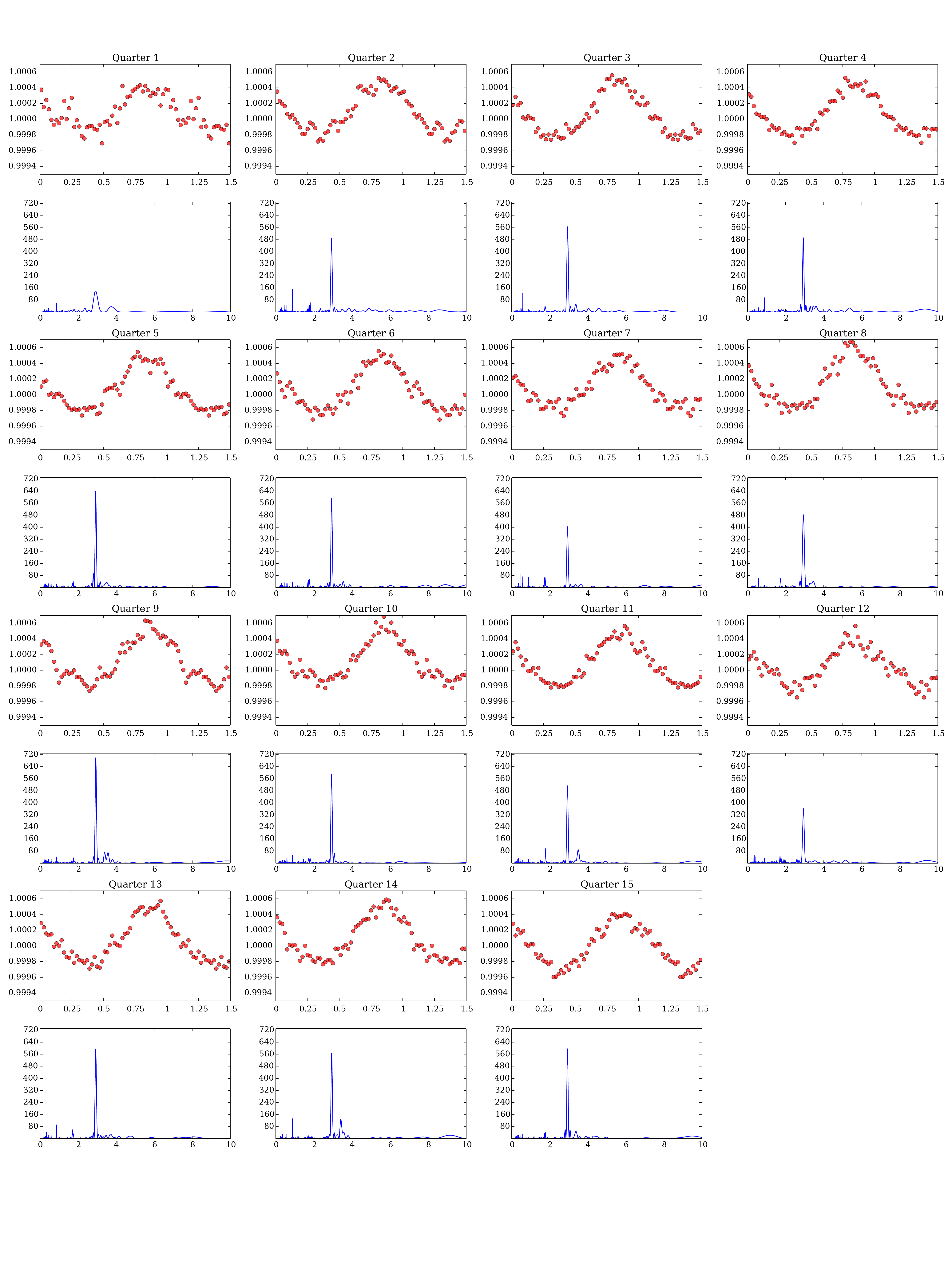}
\caption{The folded lightcurves (upper rows) and periodograms (lower rows) for the central star of J19311, for each individual quarter of data, using a period of 2.928~days. The light curves' x-axes values are phase, while those of the periodograms are time in days. The light curve fluxes are median-normalised, while the periodograms' y-axes measure the signal detection efficiency}
\label{fig:j19411-flc-multi}
\end{figure*}
\subsection{Spectroscopy}
\label{ssec:spectroscopy}

Spectra were obtained for the central star J19311 and the central star of PN Pa~5 at the Gemini North telescope with GMOS in long-slit mode using the B1200 grating between 8 and 16 July, and 3--5 September, 2012.  The spectral range was 4020--5480~\AA\ and the spectral resolution was R=4200.  Both central stars were observed in queue observing mode via phase windows defined by the photometric period from the \kepler photometry.  The central star of Pa~5 was observed in 9 out of 15 evenly spaced phase bins over {\it twice} the photometric period.  Twice the photometric period was used in the event that the photometric variability was caused by an ellipsoidal effect.  The central star J19311 was observed in 6 out of 10 evenly spaced phase bins over its photometric period.  The spectra of the central star of Pa~5 were taken in pairs of 900-second exposures while the spectra of J19311 were taken in pairs of 600-second exposures.

One spectrum of the central star of Pa~5 was also acquired at the 3.6-m WIYN telescope on Kitt Peak, on 2009, September 12, using the bench spectrograph with a resolution of R=700 and a spectral range of 4300-7000~\AA.

We also observed the central star of Kn~61 on 2012, June 16, using the National Optical Astronomy Observatory, Mayall 4-m telescope on Kitt Peak and the facility RCSpec long-slit spectrograph with the T2KA CCD. The slit was 1~arcsec wide by 49~arcsec long and oriented with a position angle of 90~deg. The KPC-22b grating in second order was used to disperse the spectra with 0.72~\AA~pixel$^{-1}$ at a nominal resolution of $\Delta \lambda$ = 1.7~\AA. The spectra covered an effective wavelength range of 4100 to 5000~\AA, being slightly out of focus at both ends where the fluxes could not be reliably calibrated. The spectrum of Kn~61 was immediately proceeded with a comparison arc lamp spectrum (FeAr) for wavelength calibration. The exposure time was 1200~sec. At least one spectrophotometric standard star was observed on the night providing a relative flux value. Calibration data consisting of bias frames, quartz lamp flat field exposures, and comparison lamp exposures were taken during the daytime.
The data reduction is based on various IRAF packages for performing image reduction and the {\it onedspec} package for extracting and calibrating the spectra. 

%%%%%%%%%%%%%%%%%%%%%%%%%%%%%%
\section{The analysis of J~19311}
\label{sec:j19311}

\subsection{The light curve of the central star J~19311}
\label{sec:lightcurvej19311}

%J19311 folded
\begin{figure}
\centering
\includegraphics[scale= 0.6]{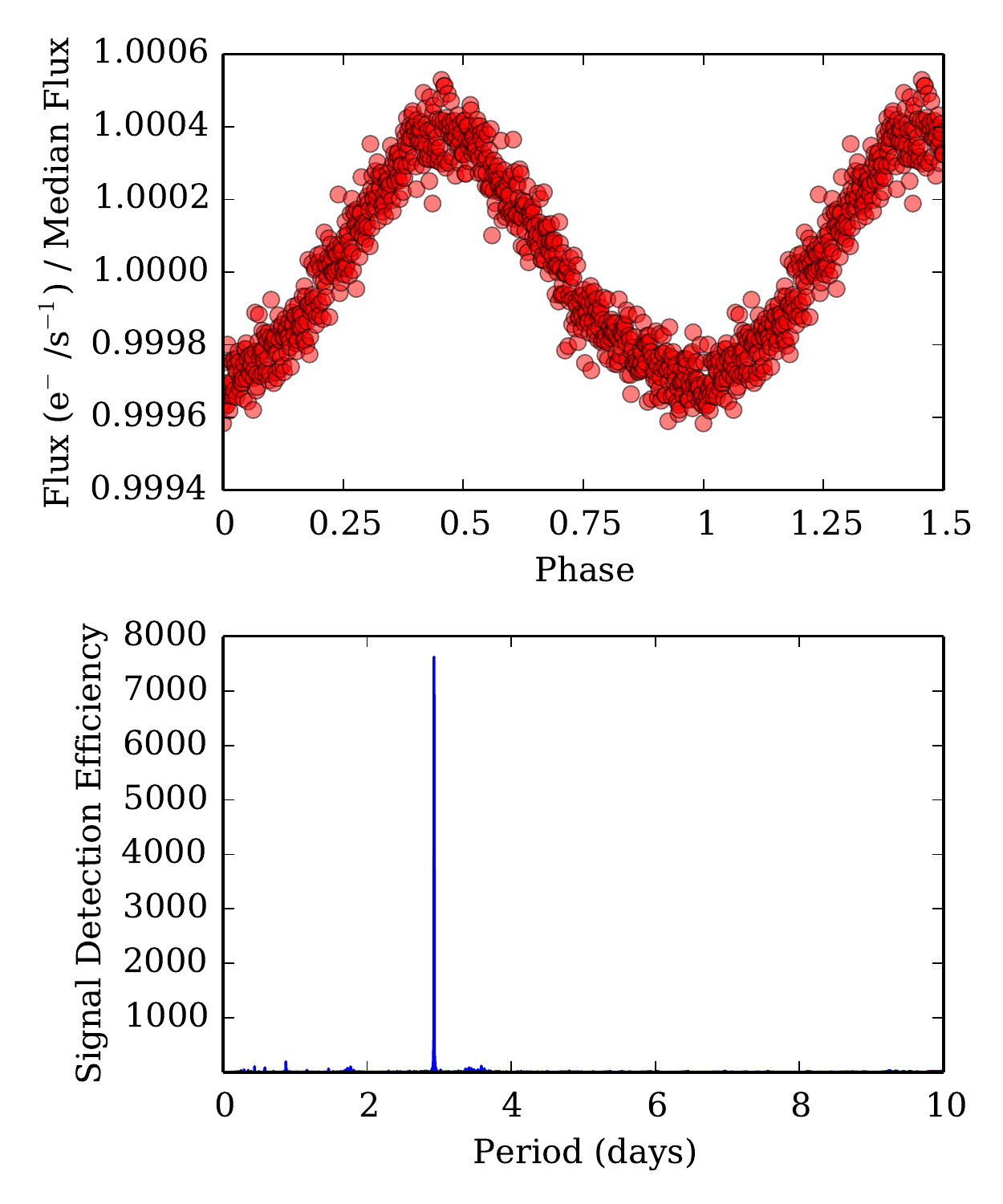}
\caption{The folded lightcurve (upper panel) and periodogram (lower panel) for the central star of J19311, for all quarters of data, using a period of 2.928 days}
\label{fig:j19411-flc-per}
\end{figure}

In Fig.~\ref{fig:j19411-flc-multi} we present the period analysis using individual quarters for the central star J19311 (the flux units in these plots are electrons per 30 minute exposure). In Fig.~\ref{fig:j19411-flc-per} we present the lightcurve phased using the most prominent period. J~19311 is almost certainly a short period binary. All quarters give consistently the same period of 2.928 days with an amplitude of 0.729 mmag. In \S~\ref{ssec:j19411-spectrum} we examine the radial velocity curve of this star together with its light curve and we will interpret this variability as due to both Doppler beaming and an ellipsoidal effect.

%In an effort to ascertain the binarity of the photometric variables we acquired time-resolved spectroscopy of two of the four variable central stars (J19311 and the central star of Pa~5; see \S~\ref{sec:sample}). We also obtained one spectrum of the central star of Kn~61. Below we describe stellar atmosphere models (either obtained from the TheoSSA database or calculated {\it ad hoc} using the T{\"u}bingen NLTE model-atmosphere package TMAP\footnote{http://astro.uni-tuebingen.de/~TMAP} \citep[][]{Werner2003,Rauch2003}) to compute non-LTE, plane-parallel, fully metal-line blanketed model atmospheres in radiative and hydrostatic  equilibrium. These models allow us to determine effective temperature and gravity parameters used as constraints for the Willson-Devinney models \citep{Wilson1971} of the variability behaviour. Averaged, rectified spectra are displayed, with some line identifications and stellar atmosphere synthetic spectral fits in Fig.~\ref{fig:spectra}.

\subsection{Modelling of the atmosphere, light and radial velocity curves of the central star J19311}
\label{ssec:j19411-spectrum}

\begin{figure*}
%\vspace{10cm}
%\includegraphics[scale= 1.0,trim=30mm 100mm 10mm 90mm,clip=true]{../Spectra/SpectralFits_NicolePlot.eps}
\includegraphics[scale= 1.0]{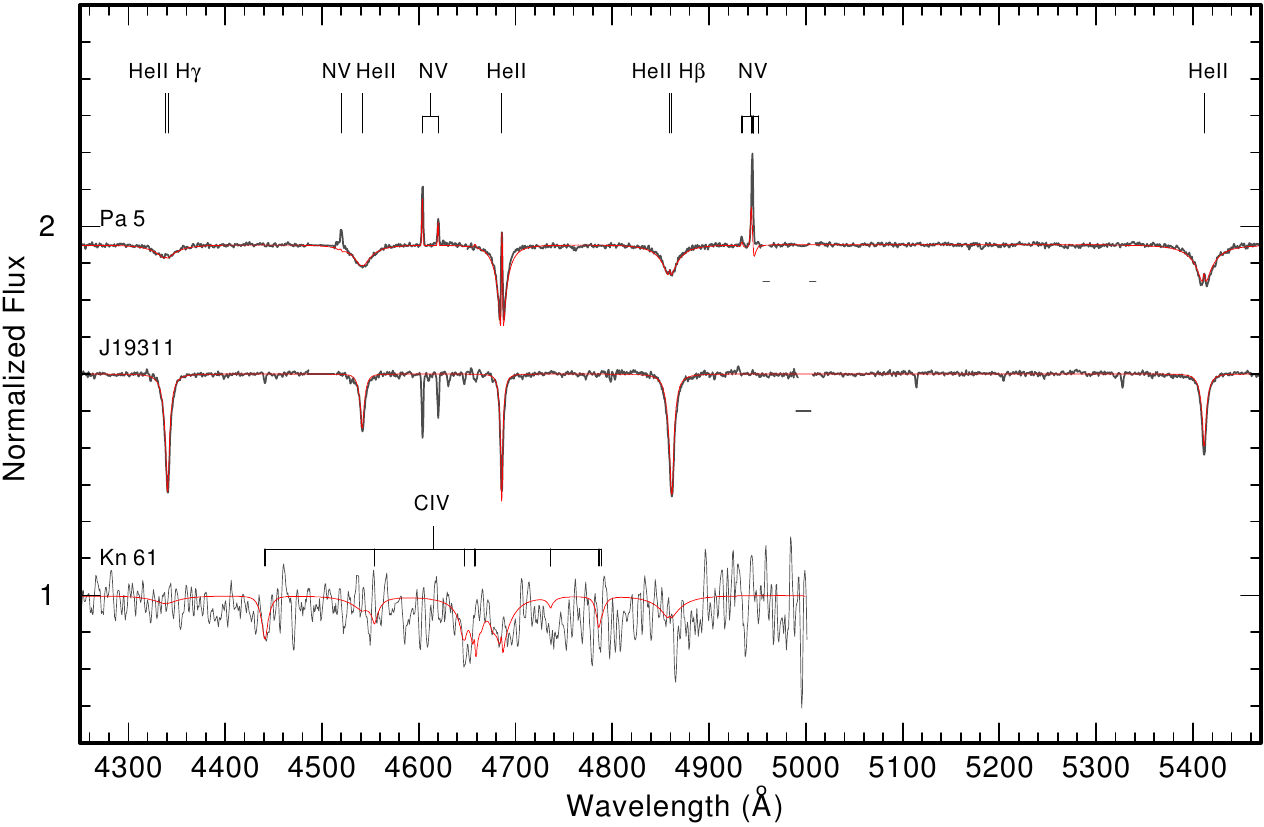}
%\special{psfile="../Spectra/SpectralFits.eps" hscale=200 vscale=200 voffset=-200}
\caption{The rectified spectra of three central stars (black lines) overlaid with the best fitting TMAP models (red lines). The spectrum of J19311 is shifted by 0.6 units, that of Pa~5's central star by 0.95}
\label{fig:spectra}
\end{figure*}

We used TMAP\footnote{astro.uni-tuebingen.de/~rauch/TMAF/flux\_H+He.html.} \citep[][]{Werner2003,Rauch2003} atmosphere models to fit the spectrum of J19311 with a pure hydrogen and helium composition, with a helium mass fraction of 0.4 which best fit the helium lines. TMAP calculates non-LTE, plane-parallel, fully metal-line blanketed model atmospheres in radiative and hydrostatic 
equilibrium. Model parameters of effective temperature and gravity, $T_{\rm CS}=80$~kK and $\log g_{\rm CS}=5.0$ fit the data well. We show the fit in Figure~\ref{fig:spectra}, alongside the fits of Pa~5 and Kn~61 which will be presented in \S~\ref{sec:kn61} and \ref{sec:pa5}. The lack of He~I lines indicates a temperature in excess of 70~kK, while the presence of NV lines in absorption (lines that we do not fit) indicates a temperature less than 115~kK. This model is only approximate but it is sufficient for a crucial constraint on the WD model below.

Within the temperature range described above we find the lower  values to be more likely, based on reported distances for the nebula.  \citet{Frew2008b} gives the distance to the PN around J19311 to be $d=1.82$ kpc.  For $T_{\rm CS}=70$~kK and our minimum radius arrived at below, we calculate $M_V = 1.98$.  Using the interstellar reddening value of \citet{Frew2008b}, $A_V=0.26$, we find the minimum distance to be $d_{min}=1.91$ kpc.  Higher temperatures will result in an intrinsically brighter central star and a greater distance required to give the same observed brightness.  If we take an uncertainty in the distance of \citet{Frew2008b} of 0.6 kpc the allowed range in temperatures is larger, but we are still limited to $T_{\rm CS}\lesssim 115$~kK.

Comparing bolometric luminosity and effective temperature values for central stars in our temperature and gravity ranges with the evolutionary models of \citet{Schoenberner1983} and \citep{Bloecker1995}, we rule out masses below $\sim$0.55 M$_\odot$.   Post-AGB stars with masses lower than this limit are intrinsically too faint.  Likewise, stars with masses above $\sim$0.85~M$_\odot$ are intrinsically too bright for these temperatures.  We therefore conclude that the central star mass is largely unconstrained , i.e., in the range  $0.55<M_{\rm CS}<0.85$ M$_\odot$.

\begin{figure}
\centering
\includegraphics[scale= 0.35,clip=true,trim=0mm 0mm 0mm 0mm]{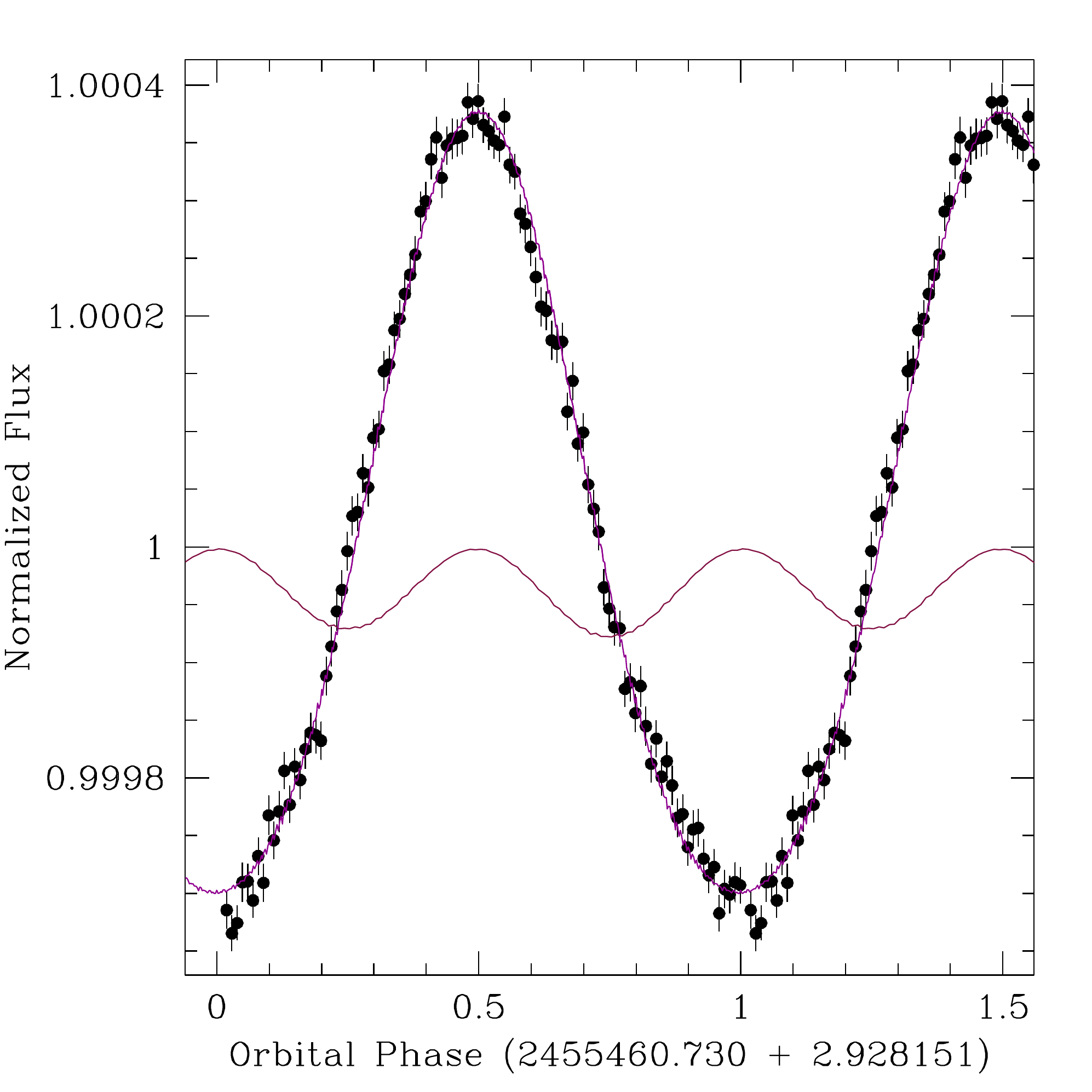}
\includegraphics[scale= 0.35,clip=true,trim=0mm 0mm 10mm 10mm]{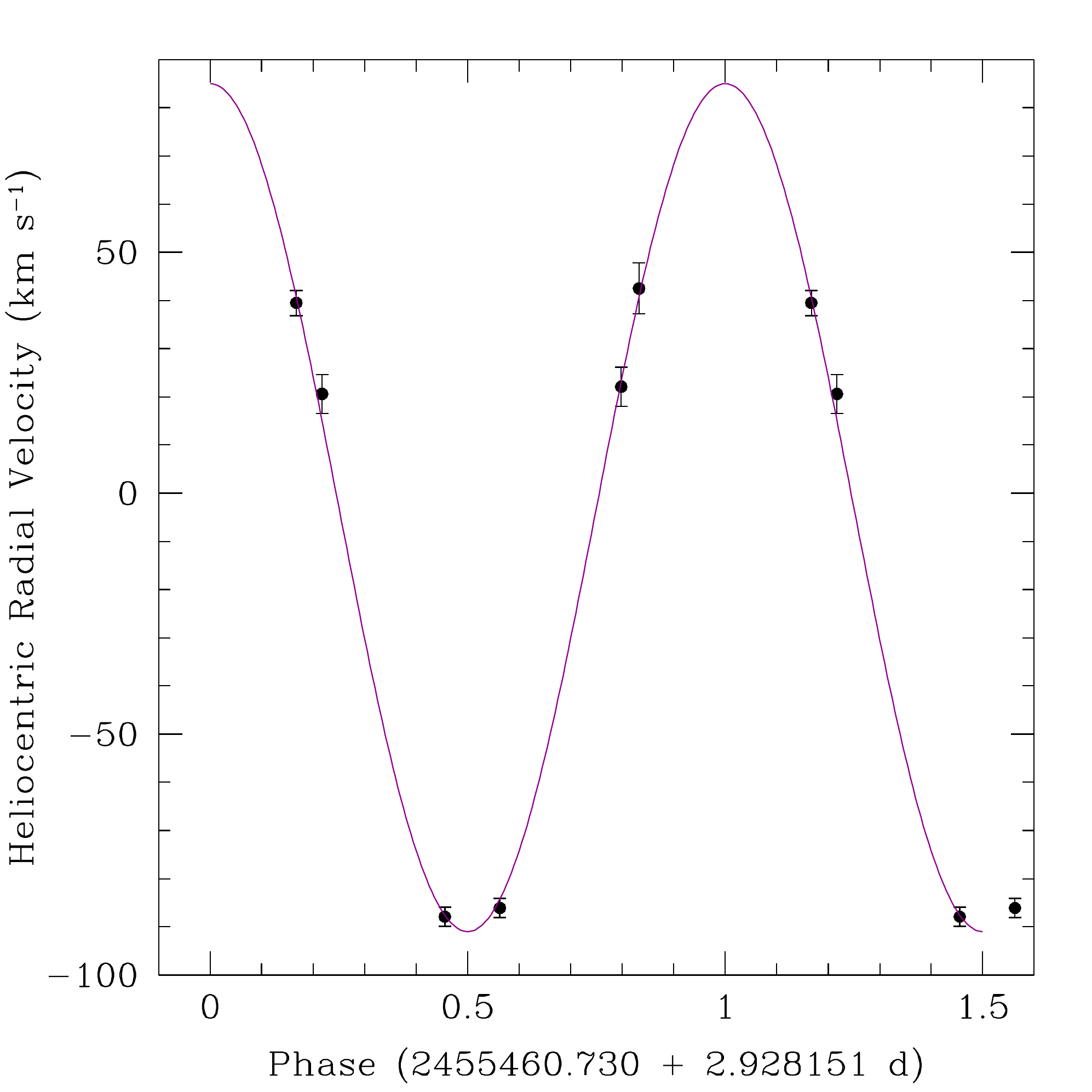}
\caption{The folded light (upper panel) and radial velocity (lower panel) curves of J19311 with overlaid the fit corresponding to the model parameters in Table~\ref{tab:WDmodel}}
\label{fig:j19411-rv}
\end{figure}

 The spectral features in J19311 show strong radial velocity variability on the same period as the \kepler photometry with a semi-amplitude of 88 km s$^{-1}$ (Figure \ref{fig:j19411-rv}).  The star is clearly in a binary system, though there is no  obvious signature of the companion star in our optical spectrum.  While the radial velocity curve is not well sampled, it is well fit by a sine curve and shows no clear sign of eccentricity in the orbit so we assume for the remainder of our analysis that the orbit is circular.  We find then that the mass function is $f(m) = (m_2 \sin i)^3/(m_1 + m_2)^2 = 0.207$~M$_\odot$, with $m_1$ and $m_2$ the masses of the primary and secondary, respectively, while $i$ is the orbital inclination.
 
The light curve for J19311 demonstrates the exquisite precision of \kepler photometry.  By coadding the photometry into 100 evenly spaced bins folded on the photometric, and orbital, period of 2.928151 days, we find a nearly sinusoidal light curve with amplitude $0.729$ mmag.  Surprisingly, the photometric light curve is out of phase with the radial velocities by exactly $\pi$. This phase shift means that the photometric variability is not due to irradiation of the companion star, in which case the phase shift would be $\pi/2$.  We can also rule out ellipsoidal variability as the primary cause of the sinusoid.  For ellipsoidal variability in which one star is distorted due to the gravity of its companion, the photometric period is half the orbital period, unlike the case here.
%Note I changed the 0.000336 normalised flux semi-amplitude into amplitude in magnitudes = 2.5xlog(1.000336)
 
It is possible that the variability is due to accretion of material from the companion, producing a cool spot on the central star.  However, it is not clear why the spot would be aligned perfectly with the axis connecting the two stars or why it would be so consistent over the duration of the \kepler observations.  There is one source of variability which we would expect to produce the observed phase shift between the radial velocities and photometry.  Relativistic effects caused by the radial motion of the star in its orbit, dominated by Doppler beaming, are known to result in photometric variability at small levels \citep{Shporer2010,Bloemen2011,Bloemen2012,Herrero2014}.
 
Using the equation for relativistic variability from \citet{Zucker2007} and \citet{Shporer2010}, with the expected temperature range, we find that our observed photometric amplitude can be described by relativistic effects given our current limits for the physical parameters of the central star.  We conclude that the system's photometric variability is due primarily to Doppler beaming.  In typical WD binaries the relativistic beaming effects of the two components are exactly out of phase and likely to nearly cancel one another out.  However, in cases such as J19311, where one of the two stars has not yet contracted to the WD cooling track and is therefore much more luminous than its companion, it dominates the system luminosity and the relativistic effects.

Given our mass function for the system and a post-AGB central star with mass $\gtrsim 0.55$ M$_\odot$, the smallest possible
companion mass, for an inclination angle, $i=90^\circ$ is $M_2(min)=0.68$ M$_\odot$.  Assuming the companion and central star to be coeval, and that the companion was less massive than the progenitor of the central star, we require that the maximum companion mass is $M_2(max)\approx8$ M$_\odot$.  Again, from our mass function we find then a minimum system inclination of $i=18^\circ$.  Using the Wilson-Devinney modeling code  \citep{Wilson1990,Wilson1971} with main sequence stars in the mass range $0.68<M_2<8.0$ M$_\odot$ and central stars with the limits we have determined above, any such companion produces an irradiation effect at least an order of magnitude larger than our observed photometric variability.  We conclude that the companion is not a main sequence star but must be a compact star, most likely a white dwarf.  If this is the case, we then also expect that the companion will have had a greater mass during the main sequence phase than the mass of the observed star's progenitor.  So we expect today's mass ratio, $q=M_2/M_{\rm CS}> 1.0$  With this requirement, a maximum WD mass of $\approx 1.4$ M$_\odot$, and our mass function derived above, we find a minimum possible inclination for the system of $i_{\rm min}=40^\circ$.  Since no eclipses are observed in the \kepler light curve, if the companion is on the WD cooling track, the inclination of the system would need to be $\lesssim 88^\circ$ for the two stars not to eclipse one another.

Alone, the relativistic variability does not provide strong constraints on the physical parameters of either star in the binary system.  The relativistic effects depend largely upon the temperatures of the two stars and their flux ratio.  With our current ranges for temperature and radius of the central star we cannot place further constraints on the binary parameters.
However, we also see that the \kepler light curve is not precisely fit by a sinusoid.  If relativistic effects are the sole cause of the light variability, the shape of the observed light curve should mimic that of the radial velocity curve, in this case a sine curve.  If we subtract a sine curve from the binned data we find that the residuals are distributed in a nearly sinusoidal curve, with a period that is half  of the orbital period and maxima that align with maxima and minima of the Doppler beaming variability.  This behaviour is what we would expect from ellipsoidal variability.  The semi-amplitude of the variability of the residuals is approximately ten times smaller than that of the variability before subtraction.

Ellipsoidal variability is caused by a difference in the projected surface area of the star, thus the amplitude of the effect depends on the geometry of the star, but not typically on the temperature.  Again we use the Wilson-Devinney code starting with our limits from above, $i>40^\circ$ and $0.55<M_{\rm CS}<0.85$ M$_\odot$ to find limits on the stellar radius (and surface gravity) that produce the observed ellipsoidal variations.  We then check the results for consistency with evolutionary models and iterate the process.  A central star with $R_{\rm CS}\approx$8--9 per cent of its Roche lobe volume radius (depending on inclination) produces the ellipsoidal variability amplitudes we observe.  The iteration process reduces the mass and radius ranges, and we find that the radius of the central star can be constrained to $0.27<R_{\rm CS}<0.35$ R$_\odot$, the surface gravity must be $5.2<\log g<5.4$, in good agreement with the preliminary model of the spectrum discussed above, and the central star mass falls in the range $0.55<M_{CS}<0.70$~M$_\odot$. In Fig.~\ref{fig:j19411-rv} we show two model curves (solid lines).  One is an example ellipsoidal variable curve, showing the amplitude of the observed ellipsoidal effect.  The second is the final model combining the Doppler beaming, as a sine curve with semi-amplitude 0.337 mmag, and the ellipsoidal effect.

Using this information for the central star and assuming the companion is either a WD or an evolved pre-WD, the relativistic variability produces our observed amplitudes best for the higher end of our temperature range.  In fact, if the companion provides less than 0.1 per cent of the total flux then the central star's temperature must be $\gtrsim 100$~kK to produce the observed amplitude. Lower central star temperatures and smaller radii within our range do work, though we find that the companion would then contribute a few percent to the system's luminosity.  The companion then must either not have reached the WD cooling track (or have expanded again possibly due to accreted material from the PN ejection), or the temperature of the companion must be very high, with $T_2\gtrsim 130$~kK.  Studies of other double-degenerate central stars, such as those of NGC~6026 \citep{Hillwig2010}, Abell~41 \citep{Shimanski2008}, and TS~01 \citep{Tovmassian2010} show companions that are either roughly coeval with the central star or have expanded and been reheated from a more evolved stage.  Since it is unlikely that all of these systems, or perhaps {\it any} of the systems, have two stars with nearly identical masses as to both be in the post-AGB/pre-WD stage simultaneously, it is more likely that compact companions to stars going through the PN stage are reheated and expand, possibly in response to accreted material (for a summary of the fit parameters see Table~\ref{tab:WDmodel}).  

\begin{table}
\begin{center}
\begin{tabular}{ll}
\hline
Parameter & Value\\
\hline
	M$_{\rm CS}$ (\msun) & 0.55 -- 0.70 \\
	M$_2$ (\msun)& 0.68 -- 1.4 \\
	R$_{\rm CS}$ (\rsun)& 0.27 -- 0.35 \\
	%R$_2$ = 0.11 \rsun\\
	T$_{\rm CS} (K)$ & 80\,000 \\
	%T$_2$ = 142\,500 K\\
	$\log (g$/cm s$^{-2}$) & 5.2 -- 5.4\\
	$i$ (deg) & 40 -- 88 \\
	$v_{\rm CS} \sin i$ (\kms)& 88\\
\hline
\end{tabular}
\caption{Model parameters to fit the light and radial velocity curves of the central star J19311. The subscript ``CS" stands for central star \label{tab:WDmodel}}
\end{center}
\end{table}

%%%%%%%%%%%%%%%%%%%%%%%%%%%%%%%%%
\section{The analysis of Kronberger 61}
\label{sec:kn61}
%Kn61 multi
\begin{figure*}
\centering
\includegraphics[scale= 0.35,clip=true,trim=0mm 0mm 0mm 0mm]{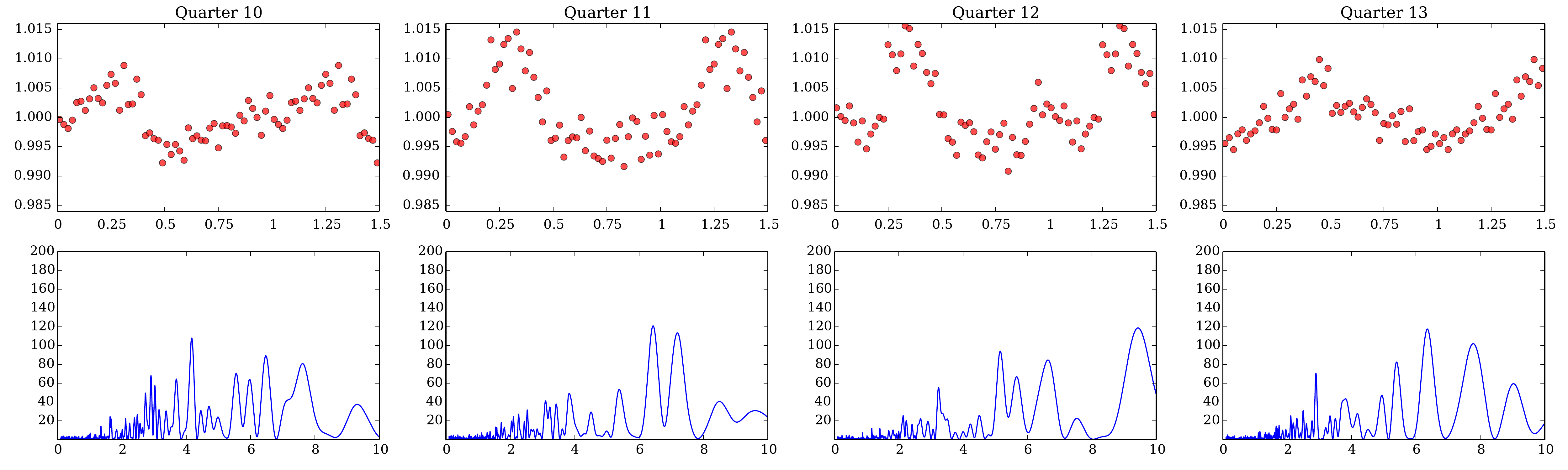}
\caption{The folded lightcurve (upper rows) and periodogram (lower rows) for the central star of Kn~61, for each individual quarter of data, using a period of 6.4~days. The light curves' x-axes values are phase, while those of the periodograms are time in days. The light curves fluxes are median-normalised, while the periodogram's y-axes measure the signal detection efficiency}
\label{fig:kn61-flc-multi}
\end{figure*}
%Kn61 folded
\begin{figure}
\centering
\includegraphics[scale= 0.6]{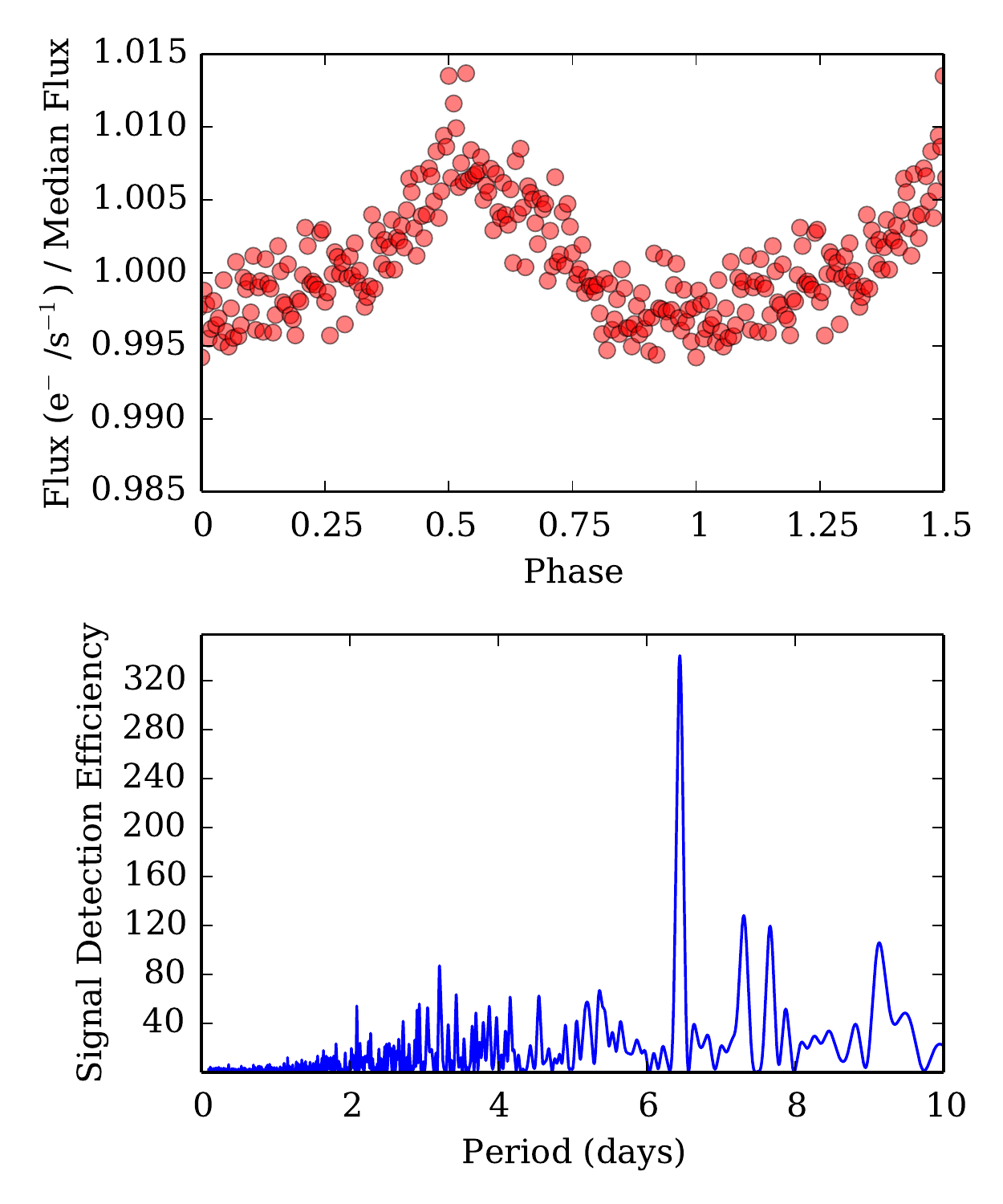}
\caption{The folded lightcurve (upper panel) and periodogram (lower panel) for the central star of Kn~61, for all quarters of data, using a period of 6.4 days}
\label{fig:kn61-flc-per}
\end{figure}
\begin{figure}
\centering
\includegraphics[scale= 0.6,clip=true,trim=0mm 40mm 0mm 10mm]{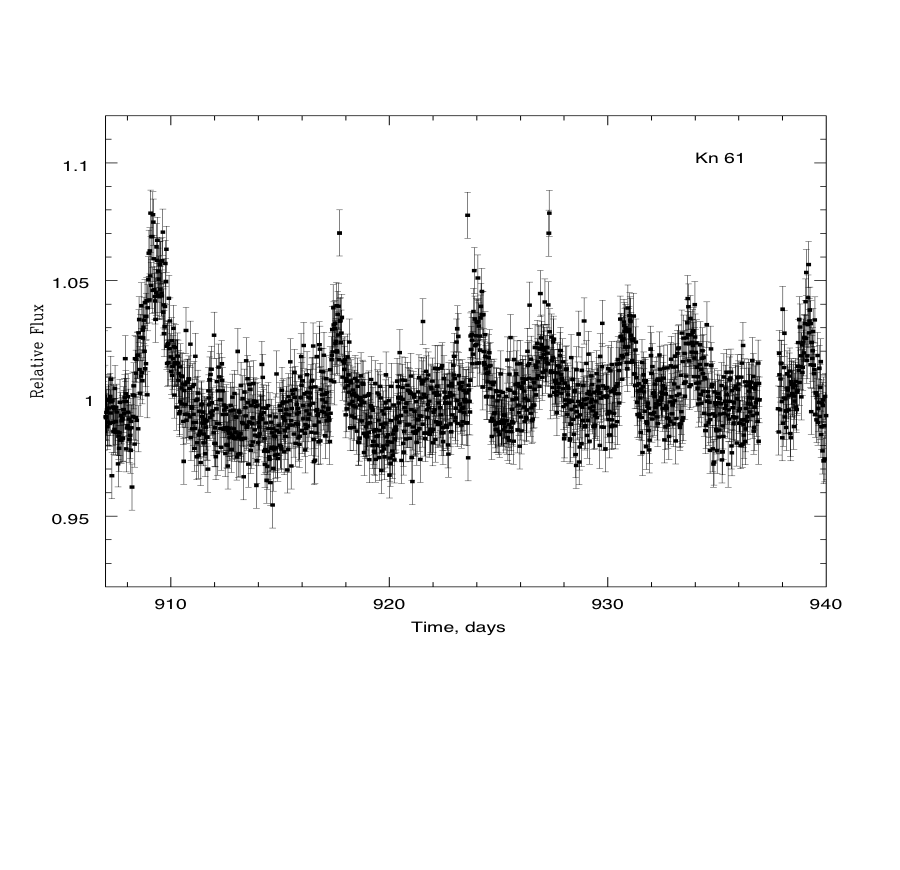}
\caption{A representative part of the lightcurve of the central star of Kn~61 showing its variability}
\label{fig:Kn61-lc}
\end{figure}
%In Fig.~\ref{fig:kn61-flc-multi} we present the period analysis using individual quarters for the central star of Kn~61 (the flux units in these plots are electrons per 30 minute exposure). In  Fig.~\ref{fig:Kn61-lc} we present a section of the lightcurve of the central star of Kn~61 (where the flux was normalised by the median flux of the quarter of data).
%N6826 multi
\begin{figure*}
\centering
\includegraphics[scale= 0.35,clip=true,trim=0mm 70mm 0mm 0mm]{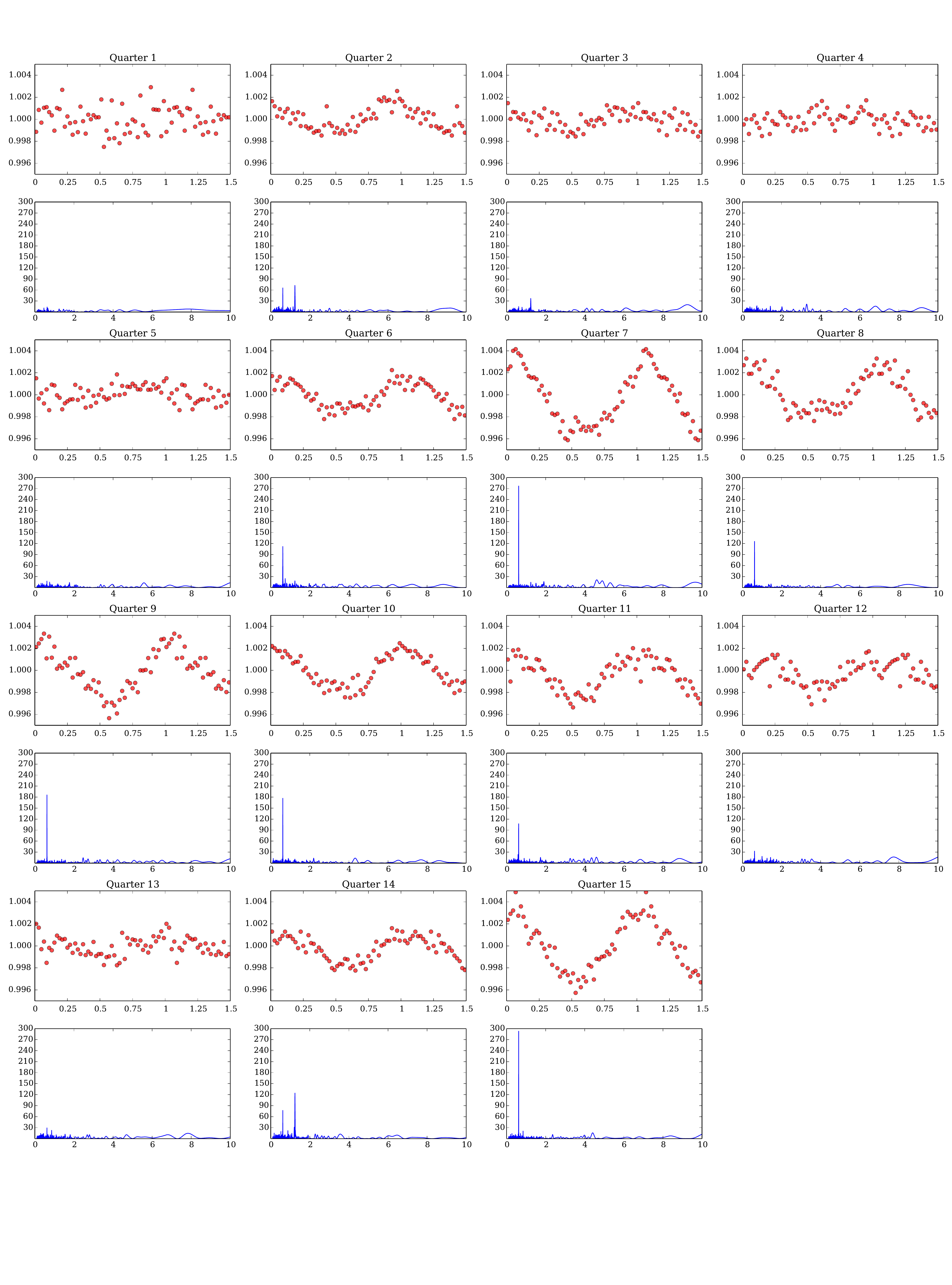}
\caption{A sample of light curves (upper rows) folded using a period of 0.61895 days, for the central star of NGC~6826, for the 11 quarters of data, with corresponding periodograms (lower rows). No consistent periodicity could be detected for this star, but see the studies of \citet{Handler2013} and \citet{Jevtic2012}. The light curves' x-axes values are phase, while those of the periodograms are time in days. The light curve fluxes are median-normalised, while the periodograms' y-axes measure the signal detection efficiency}
\label{fig:ngc6826-flc-multi}
\end{figure*}

\subsection{The light curve of the central star of Kn~61}
The central star of Kn 61 has a light curve characterised by peaks that occur with spacings of 2 to 12 days. In Fig.~\ref{fig:kn61-flc-multi} we present the period analysis using individual quarters for the central star of Kn~61 (the flux units in these plots are electrons per 30 minute exposure), while in Fig.~\ref{fig:kn61-flc-per} we have folded all quarters using a period of 6.4 days. However, inspecting the light curve (see Fig.~\ref{fig:Kn61-lc}), this cadence is clearly not any more likely than other peak recurrence times. The peaks rise from an unvarying baseline and have approximately triangular shapes, although they appear to rise faster than they decay (the smaller amplitude, shorter-duration peaks have only few data points and lower signal-to-noise ratio so it is harder to determine their shape). Brightness peaks have a width at the base between 1 and 2 days. Their height is $\sim$80-140~mmag. This type of variability is not interpretable as irradiation, ellipsoidal variability nor eclipses. In fact this variability is unprecedented and we have not been able to determine its origin. In \S~\ref{sec:conclusions} we will discuss this variability in the context of another object also discovered using \kepler data, which has similar characteristics.

\subsection{The spectrum of the central star of Kn~61 and energetics of the light outbursts}
\label{ssec:kn61-spectrum}

Our spectrum of the central star of Kn~61 shows an almost featureless continuum. However we discern the following features (see Fig.~\ref{fig:spectra}): CIV $\lambda \lambda$4440,4442,4647,4658 and HeII $\lambda$4686. H$\beta$ is absent. We modelled this low signal-to-noise spectrum using a hydrogen-free synthetic spectrum with $T_{\rm eff}$=120~kK, $\log g$=7.0 and He/C=1 (by mass). These are only indicative, but we know that the temperature must be in excess of 100~kK because \citet{GarciaDiaz2014} report to have observed the CIV lines at 5801 and 5012~\AA\ in emission. 
%PA5multi
\begin{figure*}
\centering
\includegraphics[scale= 0.3]{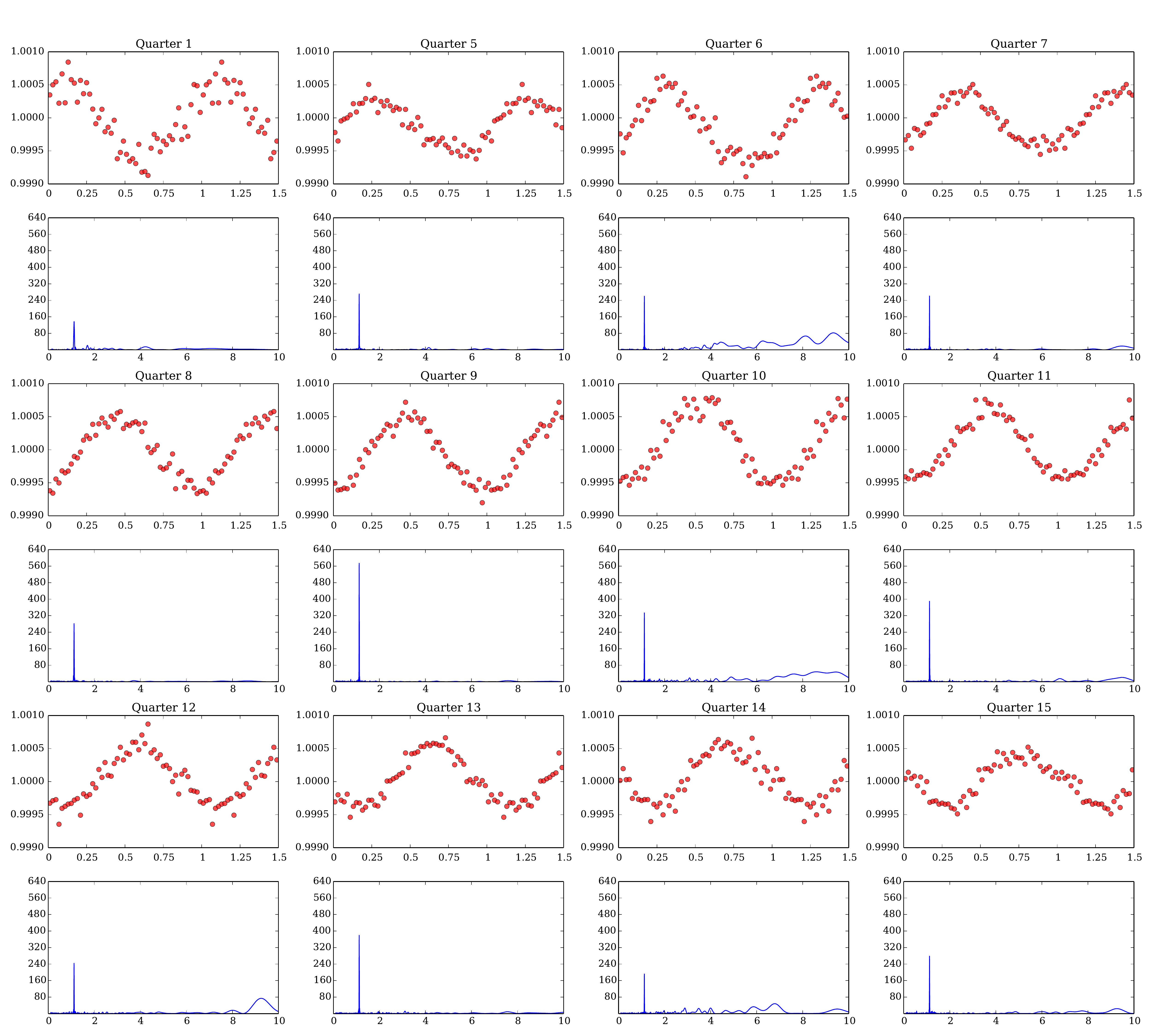}
\caption{The folded lightcurves (upper rows) and periodograms (lower rows) for the central star of Pa~5, for each individual quarter of data, using a period of 1.12 days. The light curves' x-axes values are phase, while those of the periodograms are time in days. The light curves fluxes are median-normalised, while the periodogram's y-axes measure the signal detection efficiency}
\label{fig:pa5-flc-multi}
\end{figure*}

Together with the mass-luminosity relation of \citet{Vassiliadis1994}, we obtain a radius of 0.04~\rsun, a mass of 0.5~\msun\ and a luminosity of 250~L$_\odot$. If we then interpret the light behaviour in the \kepler\ band as a similar increase in the bolometric luminosity, we derive that each peak emits 20~L$_\odot$, or 3-7$\times$10$^{39}$~ergs, assuming that the peaks are 1.1 times the quiescence brightness, have a base 1-2 days wide and have a triangular shape. If each outburst were due to an accretion event, the accreted mass would be:

\[
m = 1 \times 10^{-11} \left( {R \over {0.04\ {\rm R}_\odot}} \right) \left(E_{\rm acc} \over 10^{39}\ {\rm erg} \right) \left( M \over 0.5\ {\rm M}_\odot \right) {\rm M}_\odot
\]

\noindent We note that this calculation is quite insensitive to the value of temperature and gravity adopted. The accretion rate would be only $\sim 10^{-16}~$\msun~yr$^{-1}$ on average over a one-day-long accretion episode and these episodes would be semi-regularly spaced. 

Accretion onto the central star, however would quickly provide a hydrogen layer, which would change the PG1159 classification. Accretion onto the companion from the PG1159 star would be difficult to envisage because it would imply a compact companion and a very short periastron passage which would imply very high eccentricity. Alternative scenarios could be accretion onto the companion of material from a circumbinary disk for a light companion in an eccentric orbit. Alternatively the outbursts are caused by wind instabilities, the cause of which would require a semi-periodic nature.

%%%%%%%%%%%%%%%%%%%%%%%%%%%%%%%%%
\section{The analysis of NGC 6826}
\label{sec:ngc6826}

In Figs.~\ref{fig:ngc6826-flc-multi} we present the period analysis using individual quarters for the central star of NGC~6826 (the flux units in these plots are electrons per 30 minute exposure).  This star is peculiar. It has a strong, yet not consistent variability, exhibiting two periods: 0.619 and 1.236~days, where one is a harmonic of the other, with an amplitude varying between 2 and 8 mmag. The longer periodicity appears in only about half of the analysed quarters of data. We also measured a weak radial velocity variability  with a best period of 0.238 days (which is equivalent to one fifth of the 1.236 day periodcity), using measurements from the catalogue of \citet{Acker1982}. We discuss this peculiar behaviour further in \S~\ref{ssec:conclusionsngc6826} where we compare our measurements with those of \citet{Jevtic2012} and \citet{Handler2013}.

%This behaviour is not readily interpretable as due to binarity. This light curve has been previously studied by \citet{Jevtic2012} and \citet{Handler2013}. \citet{Handler2013} ascribe the variability to stellar rotation and derive a fast value of 73~\kms.

%

%%%%%%%%%%%%%%%%%%%%%%%%%%%%%%%%%
\section{The analysis of Patchick 5}
\label{sec:pa5}
% Pa5 folded
\begin{figure}
\centering
\includegraphics[scale= 0.6]{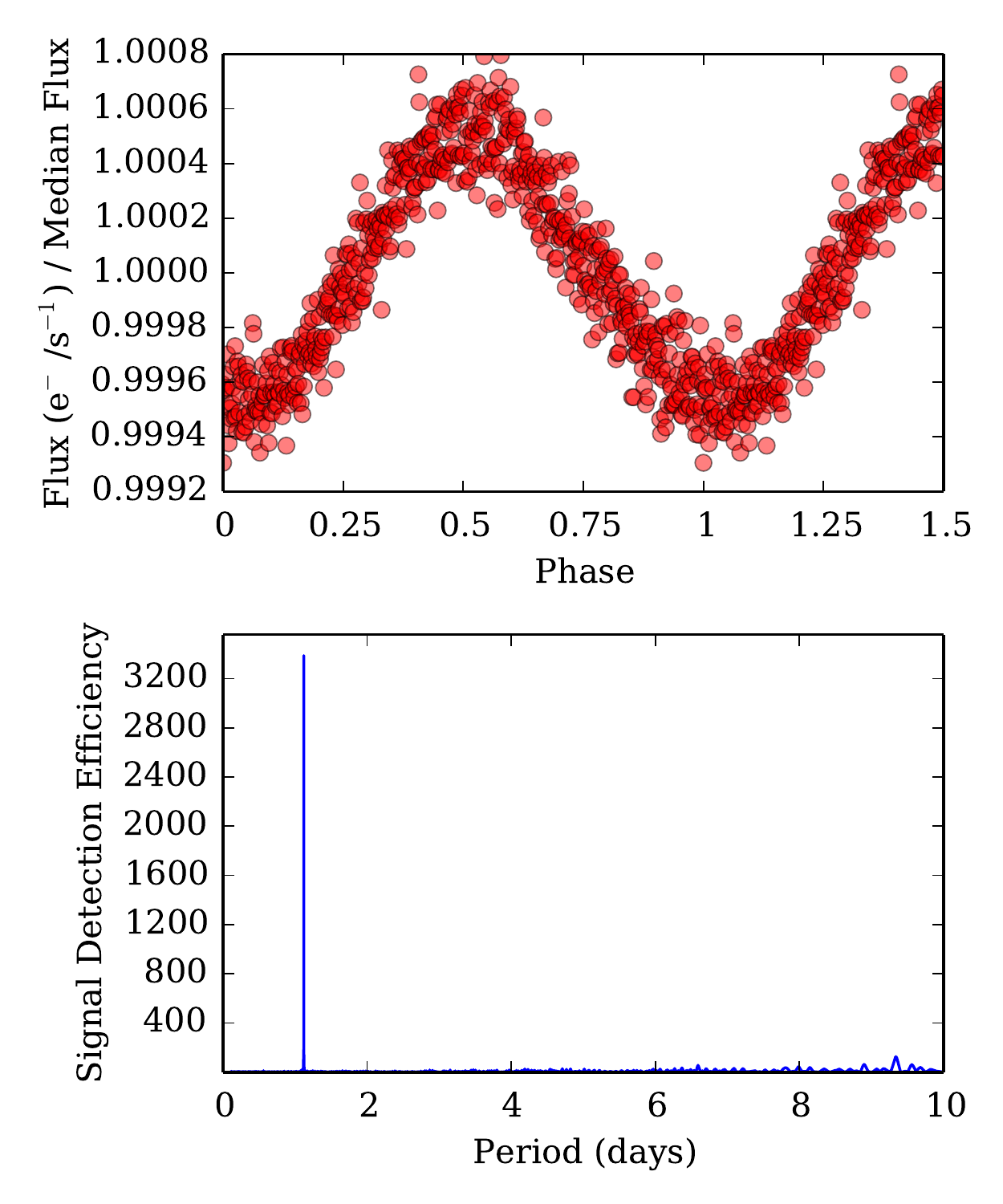}
\caption{The folded lightcurve (upper panel) and periodogram (lower panel) for the central star of Pa~5, for all quarters of data, using a period of 1.12 days}
\label{fig:pa5-flc-per}
\end{figure}
\subsection{The light and radial velocity curves of the central star of Pa~5}

In Fig.~\ref{fig:pa5-flc-multi} we present the period analysis using individual quarters for the central star of Pa~5 (the flux units in these plots are electrons per 30 minute exposure). In Fig.~\ref{fig:pa5-flc-per} we present the lightcurve phased using the most prominent period. The lightcurve of the central star of Pa~5 gives a consistent period of 1.12 days with an amplitude of 0.5~mmag. No radial velocity variability is detected larger than $\sim$5~\kms. It is entirely possible this is a binary seen close to pole on, but even so, the small amplitude of the curve implies a planetary or evolved companion. 

%The photometric variability of J19311 and the central star of Pa~5 is unlikely to have been detected from the ground. As we have discussed, percent-level differential photometry should be possible from the ground, but the poor cadence achieved, usually translates into variability limits of approximately 0.1~mag, although smaller amplitudes have been detected \citep{Bond2000,Miszalski2009}.

\subsection{Stellar atmosphere modelling of the central star of Pa~5}
\label{ssec:pa5-spectrum}

The optical spectrum of Pa\,5 strongly resembles that of the O(He) central star of PN K~1-27 \citep{Reindl2014}. The only photospheric lines in the spectrum of 
Pa\,5 are HeII lines at 4339, 4542, 4686, 4859, 5412~\AA\ and NV lines at 4513-4530, 4604, 4620, 4934 and 4943-4951~\AA\ (note that the Balmer lines are actually HeII lines and that emission lines in the trough of absorption lines are of stellar origin). 
The fact that 
the \Ionww{N}{V}{4604, 4620} lines appear in emission already indicates \Teff$\geq 115$\,kK. The central emission of 
\Ionw{He}{II}{4686} is stronger than the one observed in K\,1$-$27, suggesting that Pa\,5 is hotter than \Teffw{135}. 
To further constrain the surface parameters of Pa\,5, we calculated TMAP atmosphere models by extending the model grid calculated for K\,1$-$27 \citep{Reindl2014} up to \Teffw{160}. It included 
opacities of the elements H, He, C, N, O, and Ne, which were taken from the T{\"u}bingen model-atom database 
TMAD\footnote{http://astro.uni-tuebingen.de/~TMAD}. While we achieved a good fit for the \Ion{He}{II} and the 
\Ionww{N}{V}{4604, 4620} lines with \Teffw{145}, \loggw{6.7}, a helium mass fraction of 0.9835, N/He $= 1.3\times 10^{-2}$, C/He=5x10$^{-4}$ and O/He=4x10$^{-4}$ (by mass), we could not reproduce the 
remaining \Ion{N}{V} lines perfectly. Since the aim of our analysis was a coarse estimation of the surface parameters 
rather than a precise spectral analysis, we will leave the solution of this problem to future work. It is worth noting that, differently from the case of {K\,1$-$27}, we could not find any hint of hydrogen in the atmosphere of Pa\,5. Upper limits for the abundances  of carbon and oxygen were derived by test models where the respective lines in the model contradict the non-detection of the 
lines in the observation (at the abundance limit). For the determination of an upper limit for the carbon abundance we used 
\Ionww{C}{IV}{4441, 4646, 4658, 4660} and found C/He $\leq 5\times 10^{-4}$ (by mass) and for oxygen we used \Ionw{O}{V}{5114} and 
found O/He $\leq 4\times 10^{-4}$. Thus, Pa\,5 displays a similar CNO abundance pattern (nitrogen supersolar, carbon and oxygen subsolar, with
solar abundances according to \citet{Asplund2009}) as the O(He) stars {K\,1$-$27}, {LoTr\,4}, and 
{HS\,2209+8229} \citep{Reindl2014}.

We calculated the spectroscopic distance of Pa\,5 using the flux calibration 
of \citet{Heber1984} for $\lambda_\mathrm{eff} = 5454\,\mathrm{\AA}$,
$$ d[\mathrm{pc}]=7.11 \times 10^{4} \cdot \sqrt{H_\nu\cdot M \times 10^{0.4\, m_{\mathrm{V}_0}-\log g}} \,\, ,$$
\noindent
with $m_\mathrm{V_o} = m_\mathrm{V} - 2.175 c$, $c = 1.47 E_\mathrm{B-V}$, and 
the Eddington flux $H_\nu$ ($2.348.15\times 10^{-3}$ erg~cm$^{-2}$~s$^{-1}$~Hz${-1}$) at 5454\,\AA\,\,of our best fit model atmosphere. 
We use the visual brightness ($m_{\mathrm{V}}=15.719$~mag) obtained from the US Naval Observatory CCD Astrograph Catalogue (UCAC4, \citealt{Zacharias2013}), which is similar to the one listed in Table~1 obtained by \citet{Everett2012}.
The reddening, $E_{B-V}=0.1058$, was taken from the Galactic dust extinction maps from \cite{Schlafly2011}.
The mass of Pa\,5 (0.54\,\Msol) was derived by comparing its position in the log \Teff\ -- \logg\ plane with evolutionary tracks from 
\cite{Althaus2009}. We derive a distance of 1.4\,kpc. From the angular diameter of 2~arcmin, a linear radius of the PN of $R=0.3$\,pc 
results. Using an expansion velocity of 45~\kms \citep{GarciaDiaz2014}, the kinematic age of the PN is about 6500\,years.

Our spectrum also exhibits PN lines of [OIII] $\lambda\lambda$4959,5007. The H$\beta$ emission line in the trough of the HeII $\lambda$4859 absorption is reasonably fitted with our stellar atmosphere model, but while other HeII emission components in the trough of absorption lines are completely unchanged by the sky subtraction process, showing them to be stellar, the emission in the trough of the HeII line at 4859~\AA\ is completely removed by the sky subtraction as are the lines of [OIII], showing these to be from the nebula.

%The [OIII] and H$\beta$ emission lines are the only ones that subtract away when carrying out a background subtraction. 
%Assuming this to mean that the [OIII] lines belong to the PN we calculate a doppler shift of (29$\pm$2)~\kms from the average of the two lines. On the other hand H$\beta$ has a shift of +5~\kms. The RV discrepanc may be due to the highly uncertain H$\beta$ measurement, due to its weakeness and position in the trough of a broad absorption line. 
%The HeII and NV emission lines are not subtracted by the background subtraction process, indicating that they may come from a more compact, even unresolved structure. 
%The NV lines have shifts of 19, 13 and 3~\kms, again in disagreement with one another. These lines could derive from an irradiated atmosphere of a putative companion. Their strength remains the same over all the observations. 

The systemic velocity determined via the two [OIII] nebular lines is 26.9~\kms. \citet{GarciaDiaz2014} quote a systemic velocity value of $12.4\pm2$~\kms\ obtained from H$\alpha$ and [OIII] lines. We note that the HeII emission lines in the trough of absorption lines, as well as the H$\beta$ emission indicate a smaller systemic velocity of (9.6$\pm$1.3)~\kms, in agreement with the value of \citet{GarciaDiaz2014}. Finally the three emission lines we have identified as NV have positions corresponding with 19, 13 and 3~\kms, for the $\lambda \lambda$4604,4620 and $\lambda$4945, respectively. The nebular expansion velocity obtained from HWHM of the [OIII] lines is 76~\kms, while \citet{GarciaDiaz2014} obtained 45.2$\pm$2~\kms\ from their position-velocity diagram, a value that is more reliable and which we have adopted above. 

The WIYN spectrum of the central star of Pa~5 confirms the presence of all spectral lines. In the longer spectral range we also see H$\alpha$, both in absorption and emission (once again the absorption component is due to HeII, while the emission could be of mixed stellar and nebular origin) and the CIV doublet centred at 5806~\AA, which appears as a blue shifted absorption, with possibly a red shifted weak emission. This line was first noted by \citet{Ostensen2010}, but could not be confirmed by \citet{GarciaDiaz2014}. We also see the interstellar Na D lines, blended together. These were incorrectly identified as CaII lines by \citet{GarciaDiaz2014}.

We detected no orbital radial velocity shift to within $\pm$5~\kms. Our observations cover phases related to the photometric variability of 0.0--0.25 and 0.55--0.75, where phase zero occurs at minimum light.  If the photometric variability were caused by an irradiation effect, then the radial velocity minimum and maximum would occur at phases 0.25 and 0.75, so we are sensitive to the full radial velocity range in that case.  For an ellipsoidal effect, our radial velocities are sensitive to about two-thirds of the full radial velocity amplitude range, thus we are highly sensitive to those variations as well. If this is a close binary, its orbital rotation axis is approximately aligned with the line of sight, a geometry that would be consistent with the small amplitude of the photometric variability, despite the short period.

%David's dissertation has E(B-V) = 0.12 for Pa5.  One option, if you're feeling like spending more time on this, is to assume that the photometric period is the rotation period.  You would then have to assume a radius, but David lists Pa5 as an evolved planetary, so unless the CS mass is very low it should be pretty small.  I don't know if that will lead anywhere, but if you need a rotation rate that gives a period of less than 1.1 days then it might show the variability can't be due to rotationÉ

\subsection{Wilson-Devinney models of the photometric variability of the central stars of Pa~5}
\label{ssec:WDmodels-Pa5}

The sinusioidal variability in the \kepler light curve of the central star of Pa~5 seems to indicate the presence of a binary companion.  However, there are several post-AGB and WD stars such as WD~1953-011 \citep{Wade2003, Valyavin2011}, BOKS 53586 \citep{Holberg2011}, and WD~0103+732, (Hillwig et al., in preparation) that exhibit a consistent periodic behaviour in which the variability has been attributed to magnetic spots. In these cases the periods reflect the rotation of the star rather than an orbital modulation.  
%In fact in temperature and variability amplitude the central star of Pa~5 is very similar to WD~0103+732, though its 1.12 day period is significantly longer than for WD~0103+732. ORSOLA: this is not so, as indicated by Todd in e-mail because the full amplitude of the variation of Pa5 is 0.5 mmag, not 50 mmag.

We have explored the potential binary origins for the photometric variability in the central star of Pa~5.  We can effectively rule out most stellar companions because of the lack of radial velocity variation of the spectral lines and because of the lack of emission lines in the spectrum that would be caused by irradiation.  While we can find arbitrarily low inclinations for which the amplitude of an irradiation effect matches the observed \kepler photometry, for such a hot central star the spectrum should exhibit very strong emission lines from the irradiated stellar hemisphere.  The emission lines may not be as strong if the companion is hotter than an M or K dwarf, but then the companion's visual luminosity would approach or surpass the visual luminosity of the central star and we should see spectral features from the companion.  Therefore it is unlikely that the central star of Pa~5 has a main sequence companion in a 1.12 day orbital period.

Given the case of J19311 discussed in \S~\ref{ssec:j19411-spectrum} and the relatively large number of double-degenerate binary central stars known \citep{Hillwig2011}, it is possible that Pa~5 could have a WD or pre-WD companion.  While a cooler WD companion {\it might} exhibit an irradiation effect without strong emission lines, our limit on the radial velocity amplitude means that for a stellar mass companion the potential binary system would have an inclination $i<2.5^\circ$.

Another possibility may be a planetary mass companion.  Using the mass, radius, and temperature values above for the central star and a representative ``hot Jupiter" companion, we find that for $M_2=0.001$ M$_\odot$, $R_2=0.13$ R$_\odot$, and $T_2=1200$ K and an albedo of 0.3 for the companion, we also need to resort to a very low inclination to match the photometric amplitude, arriving at $i=1.5^\circ$.

Yet another alternative is that the variability is caused by a spot of constant size and constant  temperature contrast, at a constant location on the surface of the star.  This would almost certainly have to be due to magnetic fields.  However, the relatively narrow spectral lines do not show evidence of a strong magnetic field and the effective temperature is {\it much} too high to expect a convective atmosphere which may help produce spots \citep[for a recent review see][]{Brinkworth2013}. Weak magnetic fields ($\sim$100~G) have only been detected once in a PN central star \citep{Todt2014}, while several searches have returned either false positives \citep{Jordan2005} or upper limits of few $\times 100$~G \citep{Jordan2012,AsensioRamos2014}.

%\section{Stellar atmosphere modelling and the search for radial velocity variability }
%\label{sec:rv}
%In an effort to ascertain the binarity of the photometric variables we acquired time-resolved spectroscopy of two of the four variable central stars (J19311 and the central star of Pa~5; see \S~\ref{sec:sample}). We also obtained one spectrum of the central star of Kn~61. Below we describe stellar atmosphere models (either obtained from the TheoSSA database or calculated {\it ad hoc} using the T{\"u}bingen NLTE model-atmosphere package TMAP\footnote{http://astro.uni-tuebingen.de/~TMAP} \citep[][]{Werner2003,Rauch2003})to compute non-LTE, plane-parallel, fully metal-line blanketed model atmospheres in radiative and hydrostatic equilibrium. These models allow us to determine effective temperature and gravity parameters used as constraints for the Willson-Devinney models \citep{Wilson1971} of the variability behaviour. Averaged, rectified spectra are displayed, with some line identifications and stellar atmosphere synthetic spectral fits in Fig.~\ref{fig:spectra}.

\section{Discussion and conclusions }
\label{sec:conclusions}

Below we explore scenarios to explain the photometric and spectroscopic behaviour of our four central stars. The central star of A~61 was not observed to vary above 2~mmag, and that of NGC6742 could not be analysed at all. Below we discuss the four objects that were found to vary.

\subsection{The short period binary J19311}
\label{ssec:}
J19311 is a short period, post-common envelope (post-CE), binary where the light variability is due to Doppler beaming rather than irradiation, and to an additional, weak, ellipsoidal variability effect. This is the first time such an effect has been detected in a central star of PN, undoubtedly because its weakness would be difficult to see with the precision and sporadic cadence of ground-based observations. The companion is almost certainly an evolved star. Despite the predicted rarity of these double degenerate central stars, they appear to be rather common among the post-CE central stars of PN \citep{Hillwig2010}. The star itself is hot, but does not appear to have any other peculiarities. Finally, the PN consists of two separate ejections, close in time and with a 90 degree angle between them \citep{Aller2013}. Common envelope ejections are far from being understood \citep[e.g.,][]{Ricker2012,Passy2012,Ivanova2013}, and this object with its two degenerate stars and the peculiar nebular morphology indicating two almost simultaneous ejections at 90~deg from one another, bears witness to the complexity of the phenomenon.

\subsection{The nature of Kn~61's central star}
\label{ssec:conclusionskn61}
The central star of Kn~61 is a variable PG1159, hydrogen-deficient star, with a variability amplitude of 0.08-0.14 mag. The approximately-triangular brightness peaks recur typically every 2-12 days. The width of the peaks at their base varies between 1 and 2 days. The shape of the light curve is very different from those of irradiated or ellipsoidal variables. 

Only one other object to our knowledge may be similar to this: the DA WD J1916+3938 (KIC~4552982). Aside from pulsations \citep{Hermes2011}, it shows a \kepler light curve variability very similar to that exhibited by the central star or Kn~61, except with a slightly larger amplitude of 0.15 mag from base to peak (JJ Hermes, private communication). The peaks repeat approximately every 3 days (cf. our 2-12 days) and they are narrower, only $\sim$10 hours vs. our 24-48 hours. This star is an 11-kK WD with a $\log g = 8.3$ and as such is very different from our hot pre-WD. Interpreting the brightness peaks of J1916+3938 as accretion events the median accreted mass per episode is 5$\times 10^{-18}$~\msun, or an accretion rate of $10^{-22}$~\msun~yr$^{-1}$, over 10 hour-long events, substantially less than equivalent numbers deduced for Kn~61's central star. 

Another PG1159 star is known to have outbursts, Lo~4 \citep{Werner1992}. \citet{Bond2014} measured them to recur approximately every 100 days and to occupy approximately 5 per cent of the monitoring period. These outbursts do not appear to have any relationship with the pulsations of the star, which are of the order of minutes, nor to be caused by them. If the outbursts on Kn~61's central star are due to accretion episodes, each would accrete $\sim 10^{-11}$~\msun (assuming the accretion happens onto the central star). Accretion of, presumably, hydrogen-rich material onto a hydrogen deficient, PG1159 star would rapidly change its spectral type. 

If the light curve peaks were instead caused by accretion onto a companion orbiting the PG1159 primary, caused perhaps by periastron passages, then the companion would have to be more compact than the PG1159 star, hence a WD. However, in order to have an orbit with a period of the order of days and with periastron passages that allow a very small star to overflow its Roche lobe, the eccentricity would have to be unreasonably high. Another alternative are accretion from a circumbinary disk onto a companion that, being lighter than the central star and in an eccentric orbit, moves periodically away from the centre of mass and into the region of the disk. Alternatively these 20-\lsun\ light outbursts are semi-periodic wind outbursts, which may also have been responsible for the nebular morphology (see Fig.~2), but we then wonder what causes them to recur. It is possible that wind outbursts are triggered by a companion in an eccentric orbit, generating a mechanism similar to the so-called ``heartbeat" stars \citep{Thompson2012}. All of these scenarios involve a short-period eccentric binary having evolved through a CE interaction,but so far our understanding of CE interactions is that the eccentricity of the post-CE binary would be relatively low \citep[$\lesssim0.2$;][]{Passy2012}. 

A more in-depth analysis of the light curve should be carried out to determine whether the 6.4 day period is the real period, with additional variability complicating the power spectrum. A time-resolved spectral analysis should also help to shed light on the origin of this variability.

We have considered the possibility that a foreign, non-associated object affected the light curve. We have carefully investigated this possibility by reanalysing the {\it Kepler} data and we found no indication that this is the case. However, a chance alignment remains a possibility given {\it Kepler}'s low spatial resolution. This said, this light curve is not just odd for a central star of PN, it is unique of its own right. We therefore appeal to Occam's Razor and argue that it is unlikely that a chance alignment was observed between a rare central star of PN and a unique object with light properties that have not been observed before.

We must in the end conclude that we have at this point no definitive explanation for the light behaviour of the central star of Kn~61.

\subsection{The fast-rotating central star of NGC6826}
\label{ssec:conclusionsngc6826}
The \kepler\ data of the central star of NGC~6826 was carefully analysed by \citet{Jevtic2012} and \citet{Handler2013}. While the former study ascribes the incoherent variability to a combination of binarity and wind variability, the latter concluded that the variability is due to rotationally-induced wind variability. We have measured a distinct (2-8 mmag) dominant periodicity of 0.619 days with an occasional periodicity of 1.236 days, appearing in only about half of the analysed quarters of data. We also measured a weak radial velocity variability with a best period of 0.238 days (which is equivalent to one fifth of the 1.236 days).  \citet{Handler2013} on the other hand measured several periodicities but a persistent periodicity emerges in their data at 1.23799 days, which, as the lowest harmonic, they conclude is related to the underlying cause of the phenomenon. Their spectra show primarily non periodic variability, although their HeII lines show a period at 0.13 days. They exclude any spectral variability on the dominant photometric period of 1.24 days. 

The conclusion for this object is that the variability is not due to a short period binary. However, it may be due to a modulation of the wind caused by rotation of the star. \citet{Handler2013} concluded that a rotation rate can be inferred from the 1.24-d period and results in a rotation rate of 73~\kms\ assuming a radius of 1.8~\rsun\ measured by \citet{Kudritzki2006}. Using a projected rotation rate of the star of  $v \sin i$ = 50~\kms\ determined by \citet{Prinja2012} an inclination angle of $\sim$45~deg was derived. Such a fast rotation would be extremely high for a post-AGB star. Measurements show that white dwarfs rotate with projected rotation velocities smaller than 10~\kms\ \citep{Berger2005} with measurements from pulsating white dwarf showing even lower values \citep{Kawaler2004}. Such low rotation rates are understood today by the effects of mass-loss and angular momentum transport \citep{Suijs2008,Cantiello2014}. These central stars have slightly larger radii compared to WDs and their rotation rates should be smaller still. It is likely that this star's rotation was caused by a merger. Mergers have been proposed to explain high rotation rates and magnetic activity in post-giant stars by, e.g., \citet[][]{Nordhaus2011} and \citet{Geier2013}.

\subsection{The periodically-variable central star of Pa~5}
\label{ssec:conclusions}
The lightcurve of the central star of Pa~5 gives a consistent period of 1.12 days with an amplitude of 0.5~mmag. No radial velocity variability is detected larger than $\sim$5~\kms. It is entirely possible this is a binary seen close to pole on, but even so, the small amplitude of the curve implies a planetary or evolved companion. A planetary companion so close to the central star would imply that planetary companions can survive a common envelope phase. Despite the fact that close planets have been announced around post-giant stars \citep[e.g.,][]{Silvotti2014} we remain suspicious that survival is possible under common envelope conditions \citep{DeMarco2012}.

It is possible that the variability is due to a star spot. However, the regularity of the light curve would imply a constant spot, which is unlikely. Also, the strong magnetic fields implied are not consistent with negative searches for magnetic fields in central stars of PN \citep{Jordan2012,AsensioRamos2014,Todt2014}. 

The central star is a rare O(He) star, in other words a star made almost entirely of helium. It is only the tenth known star in the class (four more objects have recently been found;  \citealt{Werner2010}). O(He) stars are likely the descendant of the R Coronae Borealis stars considered mergers of two white dwarfs \citep{Reindl2014,Clayton2005}. However, the central star of Pa~5 seems to have too much nitrogen and too little carbon to descend from the carbon-rich R Coronae Borealis stars. If they were mergers we may expect that the rotation rate would be higher than usual for WDs. Interpreting the 1.12 days as the rotation period, the surface velocity this star would be $v$=0.4-4.5~\kms, for radii of 0.01 to 0.1~\rsun, therefore not particularly large. 
%Only another close binary star is known at this time within the PG1159 class (though with no PN), while several others are known fast rotators \citep{Koeper2001,Rauch2003}. If the period of 1.12 days is the rotation period the surface velocity this star would be a slow rotator $v$=0.4-4.5~\kms, for radii of 0.01 to 0.1~\rsun. This would be a value appropriate for most white dwarf and a lot of slower than was was determined for PG1159 stars.	

\citet{Maoz2015} analysed 14 WDs and 7 had a variability very similar to the one of the central star of Pa~5. They reviewed several scenarios, some similar to those we have considered. Their best scenario is one where accretion of debris material along the poles increases the UV opacity generating some optically-bright spots. A misalignment between the magnetic and rotational axis would bring the spots in and out of view as the star rotates. The fact that half of their stars show the phenomenon and that this is approximately the fraction of WD that show active debris disk accretion reinforces that conclusion. It would be however unlikely to find debris disks around Pa~5's central star since, contrary to the situation in WDs, its sublimation radius is much larger than the tidal disruption radius.

At this time, the most plausible hypothesis is that Pa~5 has a central star with an evolved companion in a nearly-pole-on orbit.

\subsection{General discussion on the entire sample}
\label{ssec:conclusionssample}
There is no way to obtain a measure of the binary fraction from a sample of five objects. So, where does this small {\it Kepler} sample take us? One, possibly two, of the 5 central stars with usable data are likely binaries that would not have been detected from the ground in photometric variability surveys (though J~19311 would have been detected from the ground using radial velocity variability, a technique which is not readily used in survey mode due to its time-intensive nature \citep{DeMarco2007}). The very fast rotating central star of NGC~6826 can only be explained with current theories as the product of a merger and hence as a star that has evolved via a binary channel. Finally, the enigmatic PG1159 central star of Kn~61 with its semi-periodic small outbursts remains largely a puzzle, though it is possible that this behaviour is binary-induced.

With three out of the 4 variable objects displaying amplitudes below the ground-based detectability threshold we now know that many discoveries await surveys sporting the precision of {\it Kepler}. We also suspect that low level, periodic or semi-periodic variability is often binary related. We therefore wonder 
%draw the conclusion that 
%such low photometric-amplitude variables may be 
%common 
%and 
how much larger than 15 per cent the binary fraction would have been found, had
%that if 
\kepler 
%could have 
surveyed the entire samples surveyed by \citet{Bond2000} or \citet{Miszalski2009}.
%it surely would have found a short period binary fraction that is higher than they found ($\sim$15 per cent). 

%As explained in the introduction, a 
A short period binary fraction of 15 per cent is already higher than one would expect. This statement is based on the period distribution of intermediate mass stars \citep{Duquennoy1991,Raghavan2010}, the tidal capture radii of giant stars \cite[e.g.,][]{Villaver2009,Mustill2012} and the fact that binaries that go through a CE on the RGB tend not to  ascend the AGB \citep{Heber1986,Dorman1993}, and thus never become PN central stars. The prediction is of the order of a few percent, although the exact number depends on the mean mass of the central star population: a lower mass results in a lower number of post-CE central stars, while a higher mean mass results in a higher number.
Depending on how much larger than 15 per cent this fraction becomes,
%Increasing it by even a few percent points 
%something that may be possible even likely, in light of the current results, 
%will push the central star short period binary fraction into a regime where we will have to conclude that the 
we may face a situation where a CE phase is a preferred channel to form PN.  The ``\kepler 2" mission data will include an additional $\sim$20 PN central stars that will improve the statistics on this matter.

%One additional central star in the \kepler field may have evolved through a binary channel: the very fast rotating central star of NGC~6826 is likely to be the be the product of a merger. Finally, the enigmatic PG1159 central star of Kn~61 with its semi-periodic small outbursts remains largely a puzzle.

\section*{Acknowledgments}

We wish to thank Martin Still and Thomas Barclay of NASA Ames for providing in-depth guidance of the \kepler reduction pipeline so that we could devise the data analysis pathway for our program, and Jim Shuder and Adam Block for providing images of NGC 6742 and A 61, respectively. The Gemini Observatory and Travis Rector are acknowledged for the acquisition of the image of Kn~61.
%The amateur astronomer Richard Crisp 
Boris Gaensicke, JJ Hermes, Klaus Werner and Christian Knigge are thanked for advising us on the light curve of the central star of Kn~61. Dan Maoz is thanked for sharing the content of his paper before its publication. Some of the synthetic spectra were obtained from the German Astrophysical Virtual Observatory through the TheoSSA web interface.

This research was supported in part by NASA grant NNX12AC86G, and by the Carnegie Observatories, Pasadena, CA and includes data collected by the \kepler mission. Funding for the \kepler mission is provided by the NASA Science Mission directorate. This research made use of PyKE (Still \& Barclay 2012), a software package for the reduction and analysis of \kepler data. This open source software project is developed and distributed by the NASA \kepler Guest Observer Office. This research also made use of the NASA Exoplanet Archive, which is operated by the California Institute of Technology, under contract with the National Aeronautics and Space Administration under the Exoplanet Exploration Program. This material is based in part upon work supported by the Australian Research Council Future Fellowship (OD; Grant No. FT120100452), the National Science Foundation under Grant No. AST-1109683 (TH,OD). Any opinions, findings, and conclusions or recommendations expressed in this material are those of the author(s) and do not necessarily reflect the views of the National Science Foundation. NR is supported by the German Research Foundation (DFG, grant WE 1312/41-1).

Based in part on observations obtained at the Gemini Observatory, which is operated by the 
Association of Universities for Research in Astronomy, Inc., under a cooperative agreement 
with the NSF on behalf of the Gemini partnership: the National Science Foundation 
(United States), the National Research Council (Canada), CONICYT (Chile), the Australian 
Research Council (Australia), Minist\'{e}rio da Ci\^{e}ncia, Tecnologia e Inova\c{c}\~{a}o 
(Brazil) and Ministerio de Ciencia, Tecnolog\'{i}a e Innovaci\'{o}n Productiva (Argentina).

%\appendix
%\section[]{Appenxix example}

\bibliographystyle{../../../../../../BibliographyFiles/mn2e-james}                       %% AASTeX
\bibliography{../../../../../../BibliographyFiles/bibliography}

\begin{thebibliography}{100}
\expandafter\ifx\csname natexlab\endcsname\relax\def\natexlab#1{#1}\fi

\bibitem[{{Abell}(1966)}]{Abell1966}
{Abell} G.~O., 1966, \apj, 144, 259

\bibitem[{{Acker} {et~al}\mbox{.}(1982){Acker}, {Gleizes}, {Chopinet},
  {Marcout}, {Ochsenbein}, \& {Roques}}]{Acker1982}
{Acker} A., {Gleizes} F., {Chopinet} M., {Marcout} J., {Ochsenbein} F.,
  {Roques} J.~M., 1982, Publication Speciale du Centre de Donnees Stellaires, 3

\bibitem[{{Acker} {et~al}\mbox{.}(1992){Acker}, {Marcout}, {Ochsenbein},
  {Stenholm}, \& {Tylenda}}]{Acker1992}
{Acker} A., {Marcout} J., {Ochsenbein} F., {Stenholm} B., {Tylenda} R., 1992,
  {Strasbourg - ESO catalogue of galactic planetary nebulae. Part 1; Part 2}.
  Garching: European Southern Observatory, 1992

\bibitem[{{Aller} {et~al}\mbox{.}(2013){Aller}, {Miranda}, {Ulla},
  {V{\'a}zquez}, {Guill{\'e}n}, {Olgu{\'{\i}}n}, {Rodr{\'{\i}}guez-L{\'o}pez},
  {Thejll}, {Oreiro}, {Manteiga}, \& {P{\'e}rez}}]{Aller2013}
{Aller} A. {et~al.}, 2013, \aap, 552, A25

\bibitem[{{Althaus} {et~al}\mbox{.}(2009){Althaus}, {Panei}, {Romero},
  {Rohrmann}, {C{\'o}rsico}, {Garc{\'{\i}}a-Berro}, \& {Miller
  Bertolami}}]{Althaus2009}
{Althaus} L.~G., {Panei} J.~A., {Romero} A.~D., {Rohrmann} R.~D., {C{\'o}rsico}
  A.~H., {Garc{\'{\i}}a-Berro} E., {Miller Bertolami} M.~M., 2009, \aap, 502,
  207

\bibitem[{{Asensio Ramos} {et~al}\mbox{.}(2014){Asensio Ramos},
  {Mart{\'{\i}}nez Gonz{\'a}lez}, {Manso Sainz}, {Corradi}, \&
  {Leone}}]{AsensioRamos2014}
{Asensio Ramos} A., {Mart{\'{\i}}nez Gonz{\'a}lez} M.~J., {Manso Sainz} R.,
  {Corradi} R.~L.~M., {Leone} F., 2014, \apj, 787, 111

\bibitem[{{Asplund} {et~al}\mbox{.}(2009){Asplund}, {Grevesse}, {Sauval}, \&
  {Scott}}]{Asplund2009}
{Asplund} M., {Grevesse} N., {Sauval} A.~J., {Scott} P., 2009, \araa, 47, 481

\bibitem[{{Berger} {et~al}\mbox{.}(2005){Berger}, {Koester}, {Napiwotzki},
  {Reid}, \& {Zuckerman}}]{Berger2005}
{Berger} L., {Koester} D., {Napiwotzki} R., {Reid} I.~N., {Zuckerman} B., 2005,
  \aap, 444, 565

\bibitem[{{Bl\"ocker}(1995)}]{Bloecker1995}
{Bl\"ocker} T., 1995, \aap, 299, 755

\bibitem[{{Bloemen} {et~al}\mbox{.}(2012){Bloemen}, {Marsh}, {Degroote},
  {{\O}stensen}, {P{\'a}pics}, {Aerts}, {Koester}, {G{\"a}nsicke}, {Breedt},
  {Lombaert}, {Pyrzas}, {Copperwheat}, {Exter}, {Raskin}, {Van Winckel},
  {Prins}, {Pessemier}, {Fr{\'e}mat}, {Hensberge}, {Jorissen}, \& {Van
  Eck}}]{Bloemen2012}
{Bloemen} S. {et~al.}, 2012, \mnras, 422, 2600

\bibitem[{{Bloemen} {et~al}\mbox{.}(2011){Bloemen}, {Marsh}, {{\O}stensen},
  {Charpinet}, {Fontaine}, {Degroote}, {Heber}, {Kawaler}, {Aerts}, {Green},
  {Telting}, {Brassard}, {G{\"a}nsicke}, {Handler}, {Kurtz}, {Silvotti}, {Van
  Grootel}, {Lindberg}, {Pursimo}, {Wilson}, {Gilliland}, {Kjeldsen},
  {Christensen-Dalsgaard}, {Borucki}, {Koch}, {Jenkins}, \&
  {Klaus}}]{Bloemen2011}
{Bloemen} S. {et~al.}, 2011, \mnras, 410, 1787

\bibitem[{{Bond}(2000)}]{Bond2000}
{Bond} H.~E., 2000, in ASP Conf. Ser. 199: Asymmetrical Planetary Nebulae II:
  From Origins to Microstructures, p. 115

\bibitem[{{Bond}(2014)}]{Bond2014}
{Bond} H.~E., 2014, ArXiv e-prints

\bibitem[{{Bond} \& {Ciardullo}(1990)}]{Bond1990b}
{Bond} H.~E., {Ciardullo} R., 1990, in Astronomical Society of the Pacific
  Conference Series, Vol.~11, Confrontation Between Stellar Pulsation and
  Evolution, {Cacciari} C., {Clementini} G., eds., pp. 529--533

\bibitem[{{Brinkworth} {et~al}\mbox{.}(2013){Brinkworth}, {Burleigh}, {Lawrie},
  {Marsh}, \& {Knigge}}]{Brinkworth2013}
{Brinkworth} C.~S., {Burleigh} M.~R., {Lawrie} K., {Marsh} T.~R., {Knigge} C.,
  2013, \apj, 773, 47

\bibitem[{{Bujarrabal} {et~al}\mbox{.}(2001){Bujarrabal}, {Castro-Carrizo},
  {Alcolea}, \& {S{\'a}nchez Contreras}}]{Bujarrabal2001}
{Bujarrabal} V., {Castro-Carrizo} A., {Alcolea} J., {S{\'a}nchez Contreras} C.,
  2001, \aap, 377, 868

\bibitem[{{Cantiello} {et~al}\mbox{.}(2014){Cantiello}, {Mankovich},
  {Bildsten}, {Christensen-Dalsgaard}, \& {Paxton}}]{Cantiello2014}
{Cantiello} M., {Mankovich} C., {Bildsten} L., {Christensen-Dalsgaard} J.,
  {Paxton} B., 2014, \apj, 788, 93

\bibitem[{{Clayton} {et~al}\mbox{.}(2005){Clayton}, {Herwig}, {Geballe},
  {Asplund}, {Tenenbaum}, {Engelbracht}, \& {Gordon}}]{Clayton2005}
{Clayton} G.~C., {Herwig} F., {Geballe} T.~R., {Asplund} M., {Tenenbaum} E.~D.,
  {Engelbracht} C.~W., {Gordon} K.~D., 2005, \apjl, 623, L141

\bibitem[{{De~Marco}(2009)}]{DeMarco2009b}
{De~Marco} O., 2009, \pasp, 121, 316

\bibitem[{{De~Marco} {et~al}\mbox{.}(2004){De~Marco}, {Bond}, {Harmer}, \&
  {Fleming}}]{DeMarco2004}
{De~Marco} O., {Bond} H.~E., {Harmer} D., {Fleming} A.~J., 2004, \apjl, 602,
  L93

\bibitem[{{De~Marco} {et~al}\mbox{.}(2008){De~Marco}, {Hillwig}, \&
  {Smith}}]{DeMarco2008c}
{De~Marco} O., {Hillwig} T.~C., {Smith} A.~J., 2008, \aj, 136, 323

\bibitem[{{De Marco} {et~al}\mbox{.}(2013){De Marco}, {Passy}, {Frew}, {Moe},
  \& {Jacoby}}]{DeMarco2013}
{De Marco} O., {Passy} J.-C., {Frew} D.~J., {Moe} M., {Jacoby} G.~H., 2013,
  \mnras, 428, 2118

\bibitem[{{De Marco} {et~al}\mbox{.}(2012){De Marco}, {Passy}, {Herwig},
  {Fryer}, {Mac Low}, \& {Oishi}}]{DeMarco2012}
{De Marco} O., {Passy} J.-C., {Herwig} F., {Fryer} C.~L., {Mac Low} M.-M.,
  {Oishi} J.~S., 2012, in IAU Symposium, Vol. 282, IAU Symposium, {Richards}
  M.~T., {Hubeny} I., eds., pp. 517--520

\bibitem[{{De~Marco} {et~al}\mbox{.}(2007){De~Marco}, {Wortel}, {Bond}, \&
  {Harmer}}]{DeMarco2007}
{De~Marco} O., {Wortel} S., {Bond} H.~E., {Harmer} D., 2007, in Asymmetrical
  Planetary Nebulae IV, IAC Elec. Pub.

\bibitem[{{Dorman} {et~al}\mbox{.}(1993){Dorman}, {Rood}, \&
  {O'Connell}}]{Dorman1993}
{Dorman} B., {Rood} R.~T., {O'Connell} R.~W., 1993, \apj, 419, 596

\bibitem[{{Drummond}(1980)}]{Drummond1980}
{Drummond} J.~D., 1980, PhD thesis, New Mexico Univ., Albuquerque.

\bibitem[{{Duquennoy} \& {Mayor}(1991)}]{Duquennoy1991}
{Duquennoy} A., {Mayor} M., 1991, \aap, 248, 485

\bibitem[{{Everett} {et~al}\mbox{.}(2012){Everett}, {Howell}, \&
  {Kinemuchi}}]{Everett2012}
{Everett} M.~E., {Howell} S.~B., {Kinemuchi} K., 2012, \pasp, 124, 316

\bibitem[{{Frew}(2008)}]{Frew2008b}
{Frew} D.~J., 2008, PhD thesis, Department of Physics, Macquarie University.

\bibitem[{{Garc{\'{\i}}a-D{\'{\i}}az}
  {et~al}\mbox{.}(2014){Garc{\'{\i}}a-D{\'{\i}}az}, {Gonz{\'a}lez-Buitrago},
  {L{\'o}pez}, {Zharikov}, {Tovmassian}, {Borisov}, \&
  {Valyavin}}]{GarciaDiaz2014}
{Garc{\'{\i}}a-D{\'{\i}}az} M.~T., {Gonz{\'a}lez-Buitrago} D., {L{\'o}pez}
  J.~A., {Zharikov} S., {Tovmassian} G., {Borisov} N., {Valyavin} G., 2014,
  ArXiv e-prints

\bibitem[{{Garc{\'{\i}}a-Segura} {et~al}\mbox{.}(1999){Garc{\'{\i}}a-Segura},
  {Langer}, {R{\'o}{\.z}yczka}, \& {Franco}}]{GarciaSegura1999}
{Garc{\'{\i}}a-Segura} G., {Langer} N., {R{\'o}{\.z}yczka} M., {Franco} J.,
  1999, \apj, 517, 767

\bibitem[{{Garc{\'{\i}}a-Segura} {et~al}\mbox{.}(2005){Garc{\'{\i}}a-Segura},
  {L{\'o}pez}, \& {Franco}}]{GarciaSegura2005}
{Garc{\'{\i}}a-Segura} G., {L{\'o}pez} J.~A., {Franco} J., 2005, \apj, 618, 919

\bibitem[{{Garc{\'{\i}}a-Segura} {et~al}\mbox{.}(2014){Garc{\'{\i}}a-Segura},
  {Villaver}, {Langer}, {Yoon}, \& {Manchado}}]{GarciaSegura2014}
{Garc{\'{\i}}a-Segura} G., {Villaver} E., {Langer} N., {Yoon} S.-C., {Manchado}
  A., 2014, \apj, 783, 74

\bibitem[{{Geier} {et~al}\mbox{.}(2013){Geier}, {Heber}, {Heuser}, {Classen},
  {O'Toole}, \& {Edelmann}}]{Geier2013}
{Geier} S., {Heber} U., {Heuser} C., {Classen} L., {O'Toole} S.~J., {Edelmann}
  H., 2013, \aap, 551, L4

\bibitem[{{Gonzalez Buitrago} {et~al}\mbox{.}(2014){Gonzalez Buitrago},
  {Garc{\'{\i}}a-D{\'{\i}}az}, {L{\'o}pez}, {Zharikov}, {Tovmassian},
  {Borisov}, \& {Valyavin}}]{GonzalezBuitrago2014}
{Gonzalez Buitrago} D., {Garc{\'{\i}}a-D{\'{\i}}az} M.~T., {L{\'o}pez} J.~A.,
  {Zharikov} S., {Tovmassian} G., {Borisov} N., {Valyavin} G., 2014, in
  Asymmetrical Planetary Nebulae VI conference, Proceedings of the conference
  held 4-8 November, 2013. Edited by C. Morisset, G. Delgado-Inglada and S.
  Torres-Peimbert. Online at http://www.astroscu.unam.mx/apn6/PROCEEDINGS/,
  id.34

\bibitem[{{Handler}(2003)}]{Handler2003b}
{Handler} G., 2003, in IAU Symposium, Vol. 209, Planetary Nebulae: Their
  Evolution and Role in the Universe, {Kwok} S., {Dopita} M., {Sutherland} R.,
  eds., p. 237

\bibitem[{{Handler} {et~al}\mbox{.}(2013){Handler}, {Prinja}, {Urbaneja},
  {Antoci}, {Twicken}, \& {Barclay}}]{Handler2013}
{Handler} G., {Prinja} R.~K., {Urbaneja} M.~A., {Antoci} V., {Twicken} J.~D.,
  {Barclay} T., 2013, \mnras, 430, 2923

\bibitem[{{Harrington} {et~al}\mbox{.}(1997){Harrington}, {Borkowski}, \&
  {Tsvetanov}}]{Harrington1997}
{Harrington} J.~P., {Borkowski} K.~J., {Tsvetanov} Z.~I., 1997, in IAU
  Symposium, Vol. 180, Planetary Nebulae, {Habing} H.~J., {Lamers}
  H.~J.~G.~L.~M., eds., p. 235

\bibitem[{{Heber}(1986)}]{Heber1986}
{Heber} U., 1986, \aap, 155, 33

\bibitem[{{Heber} {et~al}\mbox{.}(1984){Heber}, {Hunger}, {Jonas}, \&
  {Kudritzki}}]{Heber1984}
{Heber} U., {Hunger} K., {Jonas} G., {Kudritzki} R.~P., 1984, \aap, 130, 119

\bibitem[{{Hermes} {et~al}\mbox{.}(2011){Hermes}, {Mullally}, {{\O}stensen},
  {Williams}, {Telting}, {Southworth}, {Bloemen}, {Howell}, {Everett}, \&
  {Winget}}]{Hermes2011}
{Hermes} J.~J. {et~al.}, 2011, \apjl, 741, L16

\bibitem[{{Herrero} {et~al}\mbox{.}(2014){Herrero}, {Lanza}, {Ribas}, {Jordi},
  {Collier Cameron}, \& {Morales}}]{Herrero2014}
{Herrero} E., {Lanza} A.~F., {Ribas} I., {Jordi} C., {Collier Cameron} A.,
  {Morales} J.~C., 2014, \aap, 563, A104

\bibitem[{{Hillwig}(2011)}]{Hillwig2011}
{Hillwig} T.~C., 2011, in Asymmetric Planetary Nebulae 5 Conference

\bibitem[{{Hillwig} {et~al}\mbox{.}(2010){Hillwig}, {Bond}, {Af{\c s}ar}, \&
  {De~Marco}}]{Hillwig2010}
{Hillwig} T.~C., {Bond} H.~E., {Af{\c s}ar} M., {De~Marco} O., 2010, \aj, 140,
  319

\bibitem[{{Holberg} \& {Howell}(2011)}]{Holberg2011}
{Holberg} J.~B., {Howell} S.~B., 2011, \aj, 142, 62

\bibitem[{{Ivanova} {et~al}\mbox{.}(2013){Ivanova}, {Justham}, {Chen}, {De
  Marco}, {Fryer}, {Gaburov}, {Ge}, {Glebbeek}, {Han}, {Li}, {Lu}, {Marsh},
  {Podsiadlowski}, {Potter}, {Soker}, {Taam}, {Tauris}, {van den Heuvel}, \&
  {Webbink}}]{Ivanova2013}
{Ivanova} N. {et~al.}, 2013, \aapr, 21, 59

\bibitem[{{Jacoby} {et~al}\mbox{.}(2012){Jacoby}, {De~Marco}, {Howell}, \&
  {Kronberger}}]{Jacoby2012}
{Jacoby} G., {De~Marco} O., {Howell} S., {Kronberger} M., 2012, in American
  Astronomical Society Meeting Abstracts, Vol. 219, American Astronomical
  Society Meeting Abstracts, p. 418.02

\bibitem[{{Jacoby} {et~al}\mbox{.}(1992){Jacoby}, {Branch}, {Ciardullo},
  {Davies}, {Harris}, {Pierce}, {Pritchet}, {Tonry}, \& {Welch}}]{Jacoby1992}
{Jacoby} G.~H. {et~al.}, 1992, \pasp, 104, 599

\bibitem[{{Jacoby} {et~al}\mbox{.}(2010){Jacoby}, {Kronberger}, {Patchick},
  {Teutsch}, {Saloranta}, {Howell}, {Crisp}, {Riddle}, {Acker}, {Frew}, \&
  {Parker}}]{Jacoby2010}
{Jacoby} G.~H. {et~al.}, 2010, \pasa, 27, 156

\bibitem[{{Jevti{\'c}} {et~al}\mbox{.}(2012){Jevti{\'c}}, {Stine}, {Nilsen},
  {Schweitzer}, {Jenkins}, {Klaus}, {Lie}, \& {McCauliff}}]{Jevtic2012}
{Jevti{\'c}} N., {Stine} P., {Nilsen} W., {Schweitzer} J.~S., {Jenkins} J.~M.,
  {Klaus} T.~C., {Lie} J., {McCauliff} S., 2012, \apj, 756, 9

\bibitem[{{Jones} {et~al}\mbox{.}(2014){Jones}, {Boffin}, {Miszalski},
  {Wesson}, {Corradi}, \& {Tyndall}}]{Jones2014}
{Jones} D., {Boffin} H.~M.~J., {Miszalski} B., {Wesson} R., {Corradi} R.~L.~M.,
  {Tyndall} A.~A., 2014, \aap, 562, A89

\bibitem[{{Jordan} {et~al}\mbox{.}(2012){Jordan}, {Bagnulo}, {Werner}, \&
  {O'Toole}}]{Jordan2012}
{Jordan} S., {Bagnulo} S., {Werner} K., {O'Toole} S.~J., 2012, \aap, 542, A64

\bibitem[{{Jordan} {et~al}\mbox{.}(2005){Jordan}, {Werner}, \&
  {O'Toole}}]{Jordan2005}
{Jordan} S., {Werner} K., {O'Toole} S.~J., 2005, \aap, 432, 273

\bibitem[{{Kawaler}(2004)}]{Kawaler2004}
{Kawaler} S.~D., 2004, in IAU Symposium, Vol. 215, Stellar Rotation, {Maeder}
  A., {Eenens} P., eds., p. 561

\bibitem[{{Kronberger} {et~al}\mbox{.}(2012){Kronberger}, {Jacoby},
  {Ciardullo}, {Crisp}, {De Marco}, {Douchin}, {Frew}, {Harmer}, {Howell},
  {Howell}, {Parker}, {Patchick}, {Rector}, {Riddle}, \&
  {Teutsch}}]{Kronberger2012}
{Kronberger} M. {et~al.}, 2012, in IAU Symposium, Vol. 283, IAU Symposium, pp.
  414--415

\bibitem[{{Kronberger} {et~al}\mbox{.}(2006){Kronberger}, {Teutsch}, {Alessi},
  {Steine}, {Ferrero}, {Graczewski}, {Juchert}, {Patchick}, {Riddle},
  {Saloranta}, {Schoenball}, \& {Watson}}]{Kronberger2006}
{Kronberger} M. {et~al.}, 2006, \aap, 447, 921

\bibitem[{{Kudritzki} {et~al}\mbox{.}(2006){Kudritzki}, {Urbaneja}, \&
  {Puls}}]{Kudritzki2006}
{Kudritzki} R.~P., {Urbaneja} M.~A., {Puls} J., 2006, in IAU Symposium, Vol.
  234, Planetary Nebulae in our Galaxy and Beyond, {Barlow} M.~J., {M{\'e}ndez}
  R.~H., eds., pp. 119--126

\bibitem[{{Kukarkin} {et~al}\mbox{.}(1981){Kukarkin}, {Kholopov}, {Artiukhina},
  {Fedorovich}, {Frolov}, {Goranskij}, {Gorynya}, {Karitskaya}, {Kireeva},
  {Kukarkina}, {Kurochkin}, {Medvedeva}, {Perova}, {Ponomareva}, {Samus}, \&
  {Shugarov}}]{Kukarkin1981}
{Kukarkin} B.~V. {et~al.}, 1981, in Moscow, Acad. of Sciences USSR
  Sternberg,1951 (1981), p.~0

\bibitem[{{Lagadec} \& {Zijlstra}(2008)}]{Lagadec2008}
{Lagadec} E., {Zijlstra} A.~A., 2008, \mnras, 390, L59

\bibitem[{{Long} {et~al}\mbox{.}(2013){Long}, {Jacoby}, {De Marco}, {Hillwig},
  {Kronberger}, \& {Howell}}]{Long2013}
{Long} J., {Jacoby} G., {De Marco} O., {Hillwig} T.~C., {Kronberger} M.,
  {Howell} S.~B., 2013, in American Astronomical Society Meeting Abstracts,
  Vol. 221, American Astronomical Society Meeting Abstracts \#221, p. \#249.07

\bibitem[{{Maoz} {et~al}\mbox{.}(2015){Maoz}, {Mazeh}, \&
  {McQuillan}}]{Maoz2015}
{Maoz} D., {Mazeh} T., {McQuillan} A., 2015, \mnras, 447, 1749

\bibitem[{{Marigo} {et~al}\mbox{.}(2004){Marigo}, {Girardi}, {Weiss},
  {Groenewegen}, \& {Chiosi}}]{Marigo2004}
{Marigo} P., {Girardi} L., {Weiss} A., {Groenewegen} M.~A.~T., {Chiosi} C.,
  2004, \aap, 423, 995

\bibitem[{{Mendez} {et~al}\mbox{.}(1988){Mendez}, {Kudritzki}, {Herrero},
  {Husfeld}, \& {Groth}}]{Mendez1988a}
{Mendez} R.~H., {Kudritzki} R.~P., {Herrero} A., {Husfeld} D., {Groth} H.~G.,
  1988, \aap, 190, 113

\bibitem[{{Miszalski} {et~al}\mbox{.}(2009){Miszalski}, {Acker}, {Moffat},
  {Parker}, \& {Udalski}}]{Miszalski2009}
{Miszalski} B., {Acker} A., {Moffat} A.~F.~J., {Parker} Q.~A., {Udalski} A.,
  2009, \aap, 496, 813

\bibitem[{{Moe} \& {De~Marco}(2006)}]{Moe2006}
{Moe} M., {De~Marco} O., 2006, \apj, 650, 916

\bibitem[{{Mustill} \& {Villaver}(2012)}]{Mustill2012}
{Mustill} A.~J., {Villaver} E., 2012, \apj, 761, 121

\bibitem[{{Napiwotzki}(1999)}]{Napiwotzki1999}
{Napiwotzki} R., 1999, \aap, 350, 101

\bibitem[{{Nordhaus} {et~al}\mbox{.}(2007){Nordhaus}, {Blackman}, \&
  {Frank}}]{Nordhaus2007}
{Nordhaus} J., {Blackman} E.~G., {Frank} A., 2007, \mnras, 376, 599

\bibitem[{{Nordhaus} {et~al}\mbox{.}(2011){Nordhaus}, {Wellons}, {Spiegel},
  {Metzger}, \& {Blackman}}]{Nordhaus2011}
{Nordhaus} J., {Wellons} S., {Spiegel} D.~S., {Metzger} B.~D., {Blackman}
  E.~G., 2011, Proceedings of the National Academy of Science, 108, 3135

\bibitem[{{{\O}stensen} {et~al}\mbox{.}(2010){{\O}stensen}, {Silvotti},
  {Charpinet}, {Oreiro}, {Handler}, {Green}, {Bloemen}, {Heber},
  {G{\"a}nsicke}, {Marsh}, {Kurtz}, {Telting}, {Reed}, {Kawaler}, {Aerts},
  {Rodr{\'{\i}}guez-L{\'o}pez}, {Vu{\v c}kovi{\'c}}, {Ottosen}, {Liimets},
  {Quint}, {Van Grootel}, {Randall}, {Gilliland}, {Kjeldsen},
  {Christensen-Dalsgaard}, {Borucki}, {Koch}, \& {Quintana}}]{Ostensen2010}
{{\O}stensen} R.~H. {et~al.}, 2010, ArXiv e-prints

\bibitem[{{Parker} {et~al}\mbox{.}(2006){Parker}, {Acker}, {Frew}, {Hartley},
  {Peyaud}, {Ochsenbein}, {Phillipps}, {Russeil}, {Beaulieu}, {Cohen},
  {K{\"o}ppen}, {Miszalski}, {Morgan}, {Morris}, {Pierce}, \&
  {Vaughan}}]{Parker2006}
{Parker} Q.~A. {et~al.}, 2006, \mnras, 373, 79

\bibitem[{{Passy} {et~al}\mbox{.}(2012){Passy}, {De~Marco}, {Fryer}, {Herwig},
  {Diehl}, {Oishi}, {Mac Low}, {Bryan}, \& {Rockefeller}}]{Passy2012}
{Passy} J.-C. {et~al.}, 2012, \apj, 744, 52

\bibitem[{{Prinja} {et~al}\mbox{.}(2012){Prinja}, {Massa}, \&
  {Cantiello}}]{Prinja2012}
{Prinja} R.~K., {Massa} D.~L., {Cantiello} M., 2012, \apjl, 759, L28

\bibitem[{{Raghavan} {et~al}\mbox{.}(2010){Raghavan}, {McAlister}, {Henry},
  {Latham}, {Marcy}, {Mason}, {Gies}, {White}, \& {ten
  Brummelaar}}]{Raghavan2010}
{Raghavan} D. {et~al.}, 2010, \apjs, 190, 1

\bibitem[{{Rauch} \& {Deetjen}(2003)}]{Rauch2003}
{Rauch} T., {Deetjen} J.~L., 2003, in Astronomical Society of the Pacific
  Conference Series, Vol. 288, Stellar Atmosphere Modeling, {Hubeny} I.,
  {Mihalas} D., {Werner} K., eds., p. 103

\bibitem[{{Reindl} {et~al}\mbox{.}(2014){Reindl}, {Rauch}, {Werner}, {Kruk}, \&
  {Todt}}]{Reindl2014}
{Reindl} N., {Rauch} T., {Werner} K., {Kruk} J.~W., {Todt} H., 2014, \aap, 566,
  A116

\bibitem[{{Ricker} \& {Taam}(2012)}]{Ricker2012}
{Ricker} P.~M., {Taam} R.~E., 2012, \apj, 746, 74

\bibitem[{{Schlafly} \& {Finkbeiner}(2011)}]{Schlafly2011}
{Schlafly} E.~F., {Finkbeiner} D.~P., 2011, \apj, 737, 103

\bibitem[{{Sch\"onberner}(1983)}]{Schoenberner1983}
{Sch\"onberner} D., 1983, \apj, 272, 708

\bibitem[{{Sch{\"o}nberner} {et~al}\mbox{.}(2007){Sch{\"o}nberner}, {Jacob},
  {Steffen}, \& {Sandin}}]{Schoenberner2007}
{Sch{\"o}nberner} D., {Jacob} R., {Steffen} M., {Sandin} C., 2007, \aap, 473,
  467

\bibitem[{{Schreiber} {et~al}\mbox{.}(2009){Schreiber}, {Gaensicke},
  {Zorotovic}, {Rebassa-Mansergas}, {Nebot Gomez-Moran}, {Southworth},
  {Schwope}, {Pyrzas}, {Tappert}, \& {Schmidtobreick}}]{Schreiber2009}
{Schreiber} M.~R. {et~al.}, 2009, Journal of Physics Conference Series, 172,
  012024

\bibitem[{{Shimanskii} {et~al}\mbox{.}(2008){Shimanskii}, {Borisov},
  {Sakhibullin}, \& {Sheveleva}}]{Shimanski2008}
{Shimanskii} V.~V., {Borisov} N.~V., {Sakhibullin} N.~A., {Sheveleva} D.~V.,
  2008, Astronomy Reports, 52, 479

\bibitem[{{Shporer} {et~al}\mbox{.}(2010){Shporer}, {Kaplan}, {Steinfadt},
  {Bildsten}, {Howell}, \& {Mazeh}}]{Shporer2010}
{Shporer} A., {Kaplan} D.~L., {Steinfadt} J.~D.~R., {Bildsten} L., {Howell}
  S.~B., {Mazeh} T., 2010, \apjl, 725, L200

\bibitem[{{Silvotti} {et~al}\mbox{.}(2014){Silvotti}, {Charpinet}, {Green},
  {Fontaine}, {Telting}, {{\O}stensen}, {Van Grootel}, {Baran}, {Schuh}, \&
  {Fox Machado}}]{Silvotti2014}
{Silvotti} R. {et~al.}, 2014, \aap, 570, A130

\bibitem[{{Soker}(2006)}]{Soker2006}
{Soker} N., 2006, \pasp, 118, 260

\bibitem[{{Suijs} {et~al}\mbox{.}(2008){Suijs}, {Langer}, {Poelarends}, {Yoon},
  {Heger}, \& {Herwig}}]{Suijs2008}
{Suijs} M.~P.~L., {Langer} N., {Poelarends} A.-J., {Yoon} S.-C., {Heger} A.,
  {Herwig} F., 2008, \aap, 481, L87

\bibitem[{{Thompson} {et~al}\mbox{.}(2012){Thompson}, {Everett}, {Mullally},
  {Barclay}, {Howell}, {Still}, {Rowe}, {Christiansen}, {Kurtz}, {Hambleton},
  {Twicken}, {Ibrahim}, \& {Clarke}}]{Thompson2012}
{Thompson} S.~E. {et~al.}, 2012, \apj, 753, 86

\bibitem[{{Todt} {et~al}\mbox{.}(2014){Todt}, {Steffen}, {Hubrig},
  {Sch{\"o}ller}, {Hamann}, {Sandin}, \& {Sch{\"o}nberner}}]{Todt2014}
{Todt} H., {Steffen} M., {Hubrig} S., {Sch{\"o}ller} M., {Hamann} W.-R.,
  {Sandin} C., {Sch{\"o}nberner} D., 2014, in Asymmetrical Planetary Nebulae VI
  conference, Proceedings of the conference held 4-8 November, 2013. Edited by
  C. Morisset, G. Delgado-Inglada and S. Torres-Peimbert. Online at
  http://www.astroscu.unam.mx/apn6/PROCEEDINGS/, id.103

\bibitem[{{Tovmassian} {et~al}\mbox{.}(2010){Tovmassian}, {Yungelson}, {Rauch},
  {Suleimanov}, {Napiwotzki}, {Stasi{\'n}ska}, {Tomsick}, {Wilms}, {Morisset},
  {Pe{\~n}a}, \& {Richer}}]{Tovmassian2010}
{Tovmassian} G. {et~al.}, 2010, \apj, 714, 178

\bibitem[{{Valyavin} {et~al}\mbox{.}(2011){Valyavin}, {Antonyuk}, {Plachinda},
  {Clark}, {Wade}, {Fox Machado}, {Alvarez}, {Lopez}, {Hiriart}, {Han}, {Jeon},
  {Bagnulo}, {Zharikov}, {Zurita}, {Mujica}, {Shulyak}, \&
  {Burlakova}}]{Valyavin2011}
{Valyavin} G. {et~al.}, 2011, \apj, 734, 17

\bibitem[{{Vassiliadis} \& {Wood}(1994)}]{Vassiliadis1994}
{Vassiliadis} E., {Wood} P.~R., 1994, \apjs, 92, 125

\bibitem[{{Villaver} \& {Livio}(2009)}]{Villaver2009}
{Villaver} E., {Livio} M., 2009, \apjl, 705, L81

\bibitem[{{Wade} {et~al}\mbox{.}(2003){Wade}, {Bagnulo}, {Szeifert},
  {Brinkworth}, {Marsh}, {Landstreet}, \& {Maxted}}]{Wade2003}
{Wade} G.~A., {Bagnulo} S., {Szeifert} T., {Brinkworth} C., {Marsh} T.,
  {Landstreet} J.~D., {Maxted} P., 2003, in Astronomical Society of the Pacific
  Conference Series, Vol. 307, Solar Polarization, {Trujillo-Bueno} J.,
  {Sanchez Almeida} J., eds., p. 569

\bibitem[{{Werner} {et~al}\mbox{.}(2003){Werner}, {Deetjen}, {Dreizler},
  {Nagel}, {Rauch}, \& {Schuh}}]{Werner2003}
{Werner} K., {Deetjen} J.~L., {Dreizler} S., {Nagel} T., {Rauch} T., {Schuh}
  S.~L., 2003, in Astronomical Society of the Pacific Conference Series, Vol.
  288, Stellar Atmosphere Modeling, {Hubeny} I., {Mihalas} D., {Werner} K.,
  eds., p.~31

\bibitem[{{Werner} {et~al}\mbox{.}(1992){Werner}, {Hamann}, {Heber},
  {Napiwotzki}, {Rauch}, \& {Wessolowski}}]{Werner1992}
{Werner} K., {Hamann} W.-R., {Heber} U., {Napiwotzki} R., {Rauch} T.,
  {Wessolowski} U., 1992, \aap, 259, L69

\bibitem[{{Werner} {et~al}\mbox{.}(2014){Werner}, {Rauch}, \&
  {Kepler}}]{Werner2010}
{Werner} K., {Rauch} T., {Kepler} S.~O., 2014, \aap, 564, A53

\bibitem[{{Wilson}(1990)}]{Wilson1990}
{Wilson} R.~E., 1990, \apj, 356, 613

\bibitem[{{Wilson} \& {Devinney}(1971)}]{Wilson1971}
{Wilson} R.~E., {Devinney} E.~J., 1971, \apj, 166, 605

\bibitem[{{Zacharias} {et~al}\mbox{.}(2013){Zacharias}, {Finch}, {Girard},
  {Henden}, {Bartlett}, {Monet}, \& {Zacharias}}]{Zacharias2013}
{Zacharias} N., {Finch} C.~T., {Girard} T.~M., {Henden} A., {Bartlett} J.~L.,
  {Monet} D.~G., {Zacharias} M.~I., 2013, \aj, 145, 44

\bibitem[{{Zucker} {et~al}\mbox{.}(2007){Zucker}, {Mazeh}, \&
  {Alexander}}]{Zucker2007}
{Zucker} S., {Mazeh} T., {Alexander} T., 2007, \apj, 670, 1326

\end{thebibliography}

\end{document}